\documentclass[galley,epsfigi,usenatbib]{mn2e}
\usepackage{journal}
\usepackage{graphicx}
\usepackage{longtable}
\newcommand{\gsc}{SDSS~J134338.67+484426.6}
\newcommand{\gscc}{SDSS~J1343+4844}

\usepackage{times}
\usepackage{hyperref}
\title{Abundance analysis of \gsc; an extremely metal-poor star from the SDSS-MARVELS pre-survey}
\title[\gsc - \tyc ]
{Abundance analysis of \gsc;  an extremely metal-poor star from the MARVELS pre-survey}

\author[A. Susmitha Rani et al.]
	{A. Susmitha Rani$^{1,2}$,T. Sivarani$^{1}$,T.C. Beers$^{3}$, S. Fleming$^{4,5}$, S. Mahadevan$^{6,7,8}$, J. Ge$^{9}$\\
 $^{1}$Indian Institute of Astrophysics, Koramangala, Bangalore 560034,India\\
 susmitha@iiap.res.in,sivarani@iiap.res.in\\
$^{2}$ Joint Astronomy Programme,Indian Institute of Science, Bangalore, 560012,India \\
$^{3}$ Department of Physics and JINA Center for the Evolution of the Elements, University of Notre Dame, Notre Dame, IN 46556, USA \\
$^{4}$ Space Telescope Science Institute, 3700 San Martin Drive, Baltimore, MD, 21218 USA \\
$^{5}$ Computer Sciences Corporation, 3700 San Martin Drive, Baltimore, MD, 21218 USA \\
$^{6}$ Department  of  Astronomy  and  Astrophysics, The Pennsylvania  State  University, University Park, PA,16802, USA. \\
$^{7}$Center  for  Exoplanets  and  Habitable  Worlds,  The Pennsylvania  State  University,  University  Park, PA  16802, USA. \\
$^{8}$ The  Penn  State  Astrobiology  Research  Center,  The Pennsylvania  State  University,  University  Park, PA  16802, USA \\
$^{9}$ Department of Astronomy, University of Florida, Bryant Space Science Center, Gainesville, FL 32611-2055, USA} 

\begin{document}
	
\date{ Accepted ---;  Received ---;  in original
form --- \large \bf }
\pagerange{\pageref{firstpage}--\pageref{lastpage}} \pubyear{2016}

\maketitle
\label{firstpage}
\begin{abstract}

We present an elemental-abundance analysis of an extremely metal-poor
(EMP; [Fe/H] $< -3.0$) star, \gsc, identified during the course of the
MARVELS spectroscopic pre-survey of some 20000 stars to identify
suitable candidates for exoplanet searches. This star, with an apparent
magnitude $V = 12.14$, is the lowest metallicity star found in the
pre-survey, and is one of only $\sim$20 known EMP stars that are this
bright or brighter. Our high-resolution spectroscopic analysis shows
that this star is a subgiant with [Fe/H] = $-3.42$, having "normal" carbon
and no enhancement of neutron-capture abundances. Strontium is
under-abundant, [Sr/Fe] $ =-0.47$, but the derived lower limit on
[Sr/Ba] indicates that Sr is likely enhanced relative to Ba. This star
belongs to the sparsely populated class of $\alpha$-poor EMP stars that
exhibit low ratios of [Mg/Fe], [Si/Fe], and [Ca/Fe] compared to typical
halo stars at similar metallicity. The observed variations in radial
velocity from several epochs of (low- and high-resolution) spectroscopic
follow-up indicate that \gsc\ is a possible long-period binary. We also
discuss the abundance trends in EMP stars for $r$-process elements, and
compare with other magnesium-poor stars.

\end{abstract}


\section{Introduction}
       
Extremely metal-poor (EMP; [Fe/H] $< -3.0$) stars were formed during the
very early phases of our Galaxy's evolution, and thus carry the chemical
imprints of the first generations of supernovae that exploded in the Galaxy
\citep{Bromm2003ApJ, Kobayashi2006ApJ}. Detailed abundance analysis of
such stars constrain the nuclear processes and nucleosynthesis sites
which were prevalent during early galaxy evolution, and can be used to
explore predictions of Galactic chemical-evolution models
\citep{Hansen-Montes2014ApJ}.

EMP stars exhibit a variety of distinct abundance patterns (for detailed
reviews see \citealt{beers-christlieb2005ARA&A, Frebel2015ARAA}).
High-resolution spectroscopic surveys using 8-10 m class telescopes have
revealed a number of interesting trends in the abundance ratios of these
stars \citep{Cayrel2004,Barklem2005AA,Lai2008ApJ,Yong2013a,
Roederer2014AJ}. For example, \citet{Francois2007AA} showed that there
exists a large scatter in the [n-capture/Fe] ratios for stars with
[Fe/H] $< -2.5$, while the $\alpha$-elements and iron-peak elements
exhibit a surprisingly uniform behaviour. These authors also showed that
the light n-capture element abundances (e.g., Sr) are anti-correlated
with the abundances of heavier species such as Ba, suggesting the
operation of the so-called Light Element Primary Process (LEPP) among
some stars of extremely low metallicity \citep{Montes2007ApJ}. As these
trends were derived from samples of stars that did not exhibit other
anomalies in their abundance ratios, they may be representative of the
well-mixed ISM abundance at these metallicities. Similar results have
been observed in other well-mixed environments, such as globular-cluster
stars \citep{Otsuki2006ApJ}.

The majority of EMP stars in the halo and nearby dwarf galaxies 
exhibit low scatter in their $\alpha$-elements, although a few exhibit
abnormally low values compared to the the mean halo $\alpha$-element
abundance \citep{Carney1997AJ,King1997AJ,Lai2009ApJ, Aoki2009AA,
Tafelmeyer2010AA,Kirby2012AJ,Cohen2013ApJ}. Recently,
\citet{Caffau2013AA} identified a class of $\alpha$-poor EMP stars,
several of which exhibit low carbon abundances as well (see also 
\citealt{Cohen2013ApJ}, and references therein).

\citet{Kobayashi2014ApJ} showed that $\alpha$-poor EMP 
stars can be naturally explained by the nucleosynthesis yields of
core-collapse supernovae, i.e., 13 - 25 M$_\odot$ supernovae or
hypernovae, but pointed out that detailed abundances of additional
iron-peak elements, which are not available for most of this subset of
EMP stars, are required to better constrain the models.

According to \citet{Ivans2003ApJ}, the low $\alpha$-element 
abundances exhibited by stars in their sample are due to the the larger
contribution of SNIa yields over the pre-existing SNeII ejecta. However,
EMP stars might have formed much earlier than the onset of SNIa. So it
is less likely that the EMP stars are affected by the SNe Ia yields
unless they belong to a system where the star formation rate is very low
(and mixing is highly inhomogenous), so as to keep the gas cloud low in
metallicity until further star formation has taken place (after the
onset of SNe Ia).

In this paper we report low- and high-resolution spectroscopic follow-up
observations of a newly identified EMP star, \gsc\ (hereafter,
SDSS~J1343+4844) that may shed additional light on the nature of early
nucleosynthesis in the Galaxy, in particular for stars that do not
exhibit other abundance anomalies (such as carbon over-abundances) that
are commonly found for EMP stars. Since SDSS~J1343+4844 is bright
compared to many of the known EMP stars, particularly among the rare
$\alpha$-poor sub-class, it also presents an ideal target for future
high-resolution studies capable of deriving detailed isotopic ratios,
which can be used to better constrain the SNe models. 

\section{Observation and data reduction}

SDSS~J1343+4844 was identified as a likely metal-poor star from
SDSS-III-MARVELS (Multi-object APO Radial Velocity Exoplanet Large-area
Survey; \citealt{Ge2008ASPC}) pre-survey spectroscopy using the SDSS
legacy spectrographs \citep{Gunn2006AJ}. The pre-survey was used for
selecting suitable bright dwarfs for planet searches, in the magnitude
range 9 $<$ V $<$ 13 and with spectral types from late F to K. MARVELS
pre-survey plates were reduced using a slightly modified version
(NOCVS:v5\_3\_23) of the IDLSPEC2D pipeline. Stellar atmospheric
parameters were estimated using the n-SSPP pipeline, a modified version
of the SEGUE Stellar Parameter Pipeline (SSPP;
\citealt{LeeS2008AJ-SEGUE}, see also \citealt{tcbeers2014ApJ} for a
description of the use of the n-SSPP). Based on the pipeline procedures
and visual inspection, SDSS J1343+4844 was identified as a carbon-normal
EMP star. 

\begin{figure}
\centering
\includegraphics[width=8.5cm, height=6.5cm]{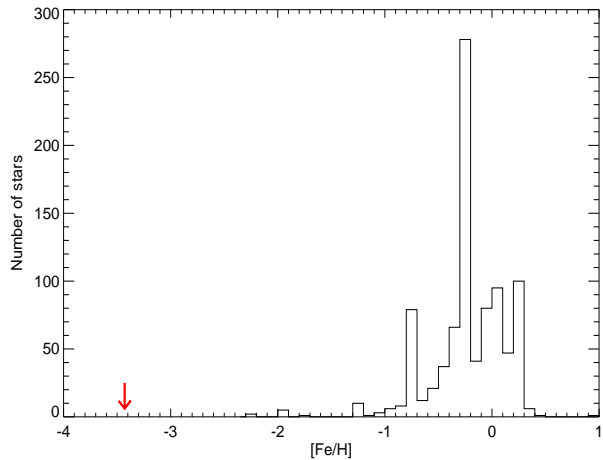}
\caption{Metallicity distribution of the MARVELS pre-survey stars.
The two strong peaks correspond to the expected metallicities
of the Galactic thin and thick disks. The arrow indicates the position
of SDSS~J1343+4844.}
\label{fig:marvels:metallicity}
\end{figure}

Figure~\ref{fig:marvels:metallicity} shows the metallicity distribution
of the complete set of MARVELS pre-survey stars; these stars clearly
exhibit peak metallicities arising from the thin-disk and thick-disk
components of the Galaxy. As is clear from inspection of this figure,
metal-poor stars are quite rare in the sample, due to the low Galactic
latitude of the pre-survey area. Among the MARVELS pre-survey
spectroscopic spectra (N $\sim$ 20000), SDSS~J1343+4844 was the only
star found with metallicity [Fe/H] $< -3.0$. 

Follow up high-resolution ($R\sim 30000$) spectroscopic observations of
SDSS~J1343+4844 were obtained with the ARC Echelle Spectrograph on the
3.5-m telescope at Apache Point Observatory (APO) and the High resolution
Spectrograph (HRS) on the Hobby-Eberly 9.2-m telescope (HET). We also
obtained multi-epoch low-resoluion ($R \sim 1300$) spectra using the
Hanle Faint Object Spectrograph Camera (HFOSC) at the 2-m Himalayan
Chandra Telescope (HCT). The spectral resolution, wavelength coverage,
and the signal-to-noise ratios of the available spectra are listed in
Table~\ref{tab:1}.

The APO pipeline-reduced high-resolution spectrum exhibited slightly
lower equivalent widths compared to the HET data. Thus, we manually
re-reduced the data using IRAF, in order to ensure proper background
subtraction, and found a better match between the equivalent widths of
the APO and HET datasets in the overlapping wavelength regions (see
Figure \ref{fig:eq_comp}). The equivalent widths for individual species
are listed in the table in the appendix. %\ref{atomic}.

\begin{figure}
\centering
\includegraphics[width=8.5cm, height=6.5cm]{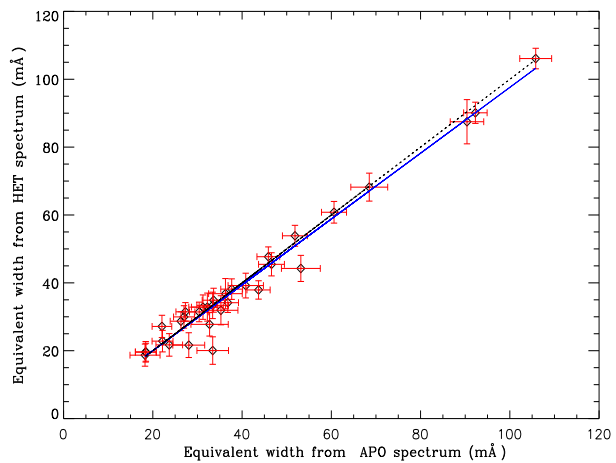}
\caption{Comparison  of measured equivalent widths of the spectral lines
from the high-resolution APO and HET spectra.
The solid line is a fit to these measurements; the dashed line is the one-to-one line.}
\label{fig:eq_comp}
\end{figure}
\begin{table*}
\centering
\caption{Details of observations and heliocentric radial velocities of \gscc\ using
various telescope/spectrograph combinations}
\begin{tabular}{c c c c c c r c}
\hline
Date          & MJD	 & Telescope-Spectrograph & Resolving Power & $\lambda$ coverage (\AA) & SNR$^{*}$ &  RV (km/s)  	\\
\hline\hline

2009-01-14  &54845.48138 & SDSS-premarvels      & 2000   & 3800 - 9200 & 58  & $-$114.7 $\pm$ 3.8  	\\
2009-03-20 & 54910.23972 & APO-ARCES 		& 31500  & 3373 - 10240& 60  & $-$123.2 $\pm$ 9.2 	\\
2012-04-23 & 56040.15483 & HET-HRS 		& 30000  & 4376 - 7838 & 58  & $-$106.1 $\pm$ 4.0	\\	
2013-06-05 & 56448.84183 & HCT-HFOSC 		& 1330   & 3800 - 6840 & 133 & $-$248.6 $\pm$ 10.4  	\\
2014-05-28 & 56805.84180 & HCT-HFOSC 		& 1330   & 3800 - 6840 & 156 & $-$219.3 $\pm$ 10.9	\\
2014-06-25 & 56833.91528 & HCT-HFOSC 		& 1330   & 3800 - 6840 & 141 & $-$241.0 $\pm$ 10.9	\\
2014-07-31 & 56869.67394 & HCT-HFOSC 		& 1330	 & 3800 - 6840 & 136 & $-$269.7 $\pm$ 10.9	\\
\hline

\end{tabular}
\begin{flushleft}
$^*$ SNR is calculated at 5000\,{\AA}.\\
\end{flushleft}
\label{tab:1}
\end{table*}

\section{Radial velocities}

Radial velocities were calculated based on six epochs of high- and
low-resolution data taken over a span of five years. A cross-correlation
analysis was performed, using the best-matching synthetic template that
represents the stellar parameters and the chemical abundances of SDSS
J1343+4844, as derived below, and degraded to match the spectral
resolution of each observation. These observed velocities were then
corrected for Earth's motion; the resultant heliocentric radial
velocities are presented in Table \ref{tab:1}.

A stability analysis of the HCT-HFOSC spectrograph was performed in
order to estimate realistic errors for the derived radial velocities of
this star. For this purpose, calibration exposures of FeAr, taken during
the course of an entire night, were used. One of the lamp exposures was
taken as a reference and relative radial velocities were derived.
Although drifts in the calibration spectra for which the telescope
position did not change are less than 2 km/sec, a large systematic shift
of the spectra was noticed for a number of different telescope
pointings; ranging from $-$30 km/s to $+$30 km/s. The systemic shift in
the spectra due to the changing telescope position has been corrected
using the prominent [OI] skylines at 5577.34, 6300.3, 6363.8 \AA\ . We
took the median of the shifts calculated from these lines and corrected
all the spectra observed with HCT accordingly. 
 
\begin{figure}
\centering
\includegraphics[width=8.5cm,height=6.5cm]{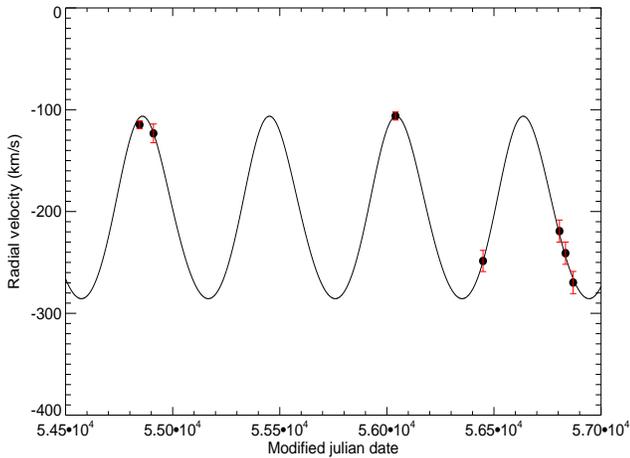}
\caption{Variation of radial velocity with MJD. The black dots represent
the observed points; the black line indicates the best-fit orbit, corresponding to a period of 592 days.   }
\label{fig:rv}
\end{figure}

We made use of the software package RVLIN, provided by
\citet{Wright2009ApjS}, which is a set of IDL routines to fit Keplerian
orbits to the radial velocity data, and find the period of the
radial-velocity variation. The radial-velocity measurements exhibit a
variation with a period of 592 days; the best fit for is shown in
Figure~\ref{fig:rv}. Although the data are insufficient to derive the
mass ratio of the binary system, the binary nature appears to be clear.
Additional radial-velocity monitoring of this system would be useful to
check on the derived period, given the sparse coverage of the proposed
orbit at present.

\section{Stellar parameters}

Low- and high-resolution spectroscopic data were both used for deriving
estimates of the effective temperature, log $\textit{g}$, and
metallicity, [Fe/H], for \gscc, as described below, so that these
determinations might be compared with one another.  

Stellar temperature estimates were derived making use, in part, of
various available photometric datasets. The photometric colours and
errors are listed in Table \ref{tab:4}. A reddening estimation (E(B-V) =
0.009) from the Schlegel et al. dust maps \citep{Schlegel1998ApJ} was
used. Optical (APASS; \citealt{Henden2015AAS}) and NIR (2MASS;
\citealt{Cutri2003}) photometric data were used to derive effective
temperature estimates, based on the Alonso relations \citep{Alonso1996,
Alonso1999} assuming a dwarf classification for a metallicity [Fe/H] =
$-3.0$ (Table~\ref{tab:phot}). 

We also used VOSA (http://svo2.cab.inta-csic.es/), the online SED fitter
to derive the temperature, using all of the available photometry
(optical, 2MASS, and WISE, \citealt{Wright2010AJ}). A Bayesian fit using
the ATLAS-NOVER-NEWODF model yielded a best-fit value of T$_{eff}$ =
5750 K. 

\begin{figure}
\centering
\includegraphics[width=8.5cm,height=6.5cm]{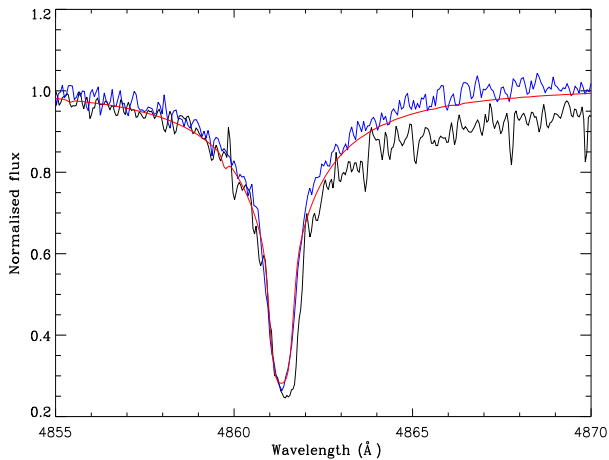}
\caption{The H$_{\beta}$ region from the APO spectrum (black), the HET
spectrum (violet), and the synthesized spectrum (red)
are shown.  The best fit corresponds to $T_{eff}$ = 5620~K.
The red wing of the APO spectrum falls at the edge of the echelle
orders, so only the blue wing 
is considered for the fit.}
\label{fig:2}
\end{figure}

We also fit the H$_{\beta}$ line wings of the APO and HET spectra for
various temperatures. The red wing of H$_{\beta}$ in the APO spectrum
lies close to the edge of an echelle order, so it was not considered in
the fitting. The best-fit synthetic spectrum in this region, along with
the observed spectra from APO and HET, are shown in Figure~\ref{fig:2}.

The derived temperature from the n-SSPP pipeline is 6161~K, which
clearly deviates from the rest of the estimates. This is due, in part,
to the fact that the flux calibration of the MARVELS pre-selection
spectrum is not accurate, as these plug-plates were observed during
twilight at high airmass. So, we re-derived T$_{eff}$ using only the
normalized SDSS spectrum, obtaining T$_{eff}$ = 5832~K. An effective
temperature of 5620~K was derived using the high-resolution spectra,
employing the excitation equilibrium for the derived abundances using Fe
I lines.

Finally, T$_{eff}$ is also derived from the V-K colour, where the
K$_{s}$ magnitude is used instead of the K magnitude (the difference
between the two magnitudes is K$_{s} - $ K = $-0.0003$, which yields no
difference in the derived temperature); we obtained T$_{eff}$ = 5671~K.
This estimate differs from the temperature obtained based on the
excitation equillibrium approach by only 50~K, which we adopt for our
subsequent analysis.

\begin{table*}
\caption{Photometric data for \gscc}
\centering
\begin{tabular}{cccccccccc}
\hline
	& B	& V	& J	& H	& K$_{s}$ & W1	&W2	&W3	&W4	\\
\hline\hline
Magnitude	&12.685 &12.147	&11.031	&10.675	&10.609	&10.566	&10.563	&10.56	&8.808	\\
error		&0.041	&0.039	&0.022	&0.028	&0.022& 0.023	&0.020	&0.081	& $\dots$ $^*$ \\
\hline
\end{tabular}
\label{tab:phot}
\begin{flushleft}
$^*$ The error is not available in the catalog.
\end{flushleft}
 
\end{table*}

\begin{table}
\caption{The temperature derived from various methods}
\centering
\begin{tabular}{c c c}
\hline
Colour/Technique	 &$ T_{eff}$ (K)	\\%& $T_{eff}$ (K)   \\
         &   (for Dwarf)        \\%& (for Giant)    \\
\hline\hline
B$-$V = 0.538 $\pm $ 0.057   	& 5618 $\pm $ 130 	\\%& 7998 $\pm $ 167	\\
V$-$K = 1.538 $\pm $ 0.044   	& 5671 $\pm $ 37	\\%& 5621 $\pm $ 40	\\
J$-$H = 0.356 $\pm $ 0.036	& 5444 $\pm $ 144  	\\%& 7990 $\pm $ 170	\\
J$-$K = 0.422 $\pm $ 0.031  	& 5394 $\pm $ 144   	\\%& 7818 $\pm $ 125	\\
SED fitting                     & 5750 &\\
SSPP                            & 6161 &\\
SDSS normalized spectra                 & 5832 &\\
High resolution spectra - Excitation equilibrium                 & 5620 &\\

\hline

\end{tabular}
\label{tab:4}
\end{table}

\begin{figure}
\centering
\includegraphics[width=8.5cm, height=6.5cm]{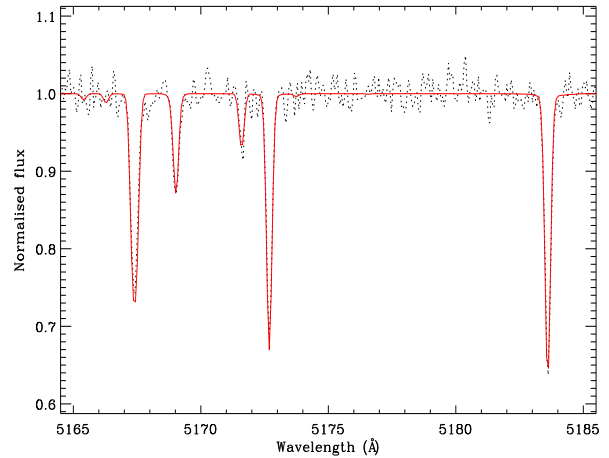}
\caption{High-resolution HET spectrum in the region of the Mg~I triplet.
The black dashed line indicates the HET spectrum, while  the red solid line indicates
the synthetic spectrum with parameters corresponding to $T_{eff}$ =
5620~K, log \textit{g} = 3.4. and [Fe/H] = $-3.42$.}
\label{fig:4}
\end{figure}

We employ several different methods for the estimation of surface
gravity of \gscc, including the ionization equilibrium of Fe~I and Fe~II
lines, isochrone fitting, and the best-fit of the Mg~I wings. We
detected only four Fe~II lines in the spectra, %(see upper panel ofFigure~\ref{fig:6})
and derived log $\textit{g}$ = 3.44. Theoretical
isochrones \citep{Demarque2004ApJS} for an age range of 11-13 Gyr have
been fitted by assuming a temperature of 5620~K, [Fe/H] = $-3.5$ and
[$\alpha$/Fe] = 0. The log $\textit{g}$ values are taken where the
temperature lines intersects the isochrones. There are two possibilities
-- the star could be a dwarf (log $\textit{g}$ $\sim$ 4.7) or a subgiant
(log $\textit{g}$ $\sim$ 3.4). We also fit the Mg~I lines for both the
APO and HET spectra at $\lambda$5167 \AA, $\lambda$5172 \AA, and
$\lambda$5183 \AA\ for various values of log $\textit{g}$, and obtained
a value of log $\textit{g}$ = 3.4 (Figure~\ref{fig:4}). We adopt a value
of log $\textit{g}$ = 3.4, a value that is consistent with all three
methods.

The microturbulent velocity ($\xi$) of the star is derived using 60 Fe~I
lines, by adjusting the input microturbulence value in such a way that
the weak and strong Fe~I lines give the same abundances.
%(see lower panel of Figure~\ref{fig:6}). 
The adopted stellar atmospheric parameters are listed in
Table \ref{tab:5}.

%\begin{figure}
%\centering
%\includegraphics[width=8.5cm, height=6.5cm]{gsc_ep_abund_review_modline.ps}
%\caption{The upper panel indicates the abundance trend for $T_{eff}$ =
%5620~K, log $\textit{g }$=3.44, $\xi$ =1.45 km/s. The red crosses
%indicate Fe~II lines, while the black crosses are Fe~I lines. The lower
%panel indicates the derived Fe abundance as a function of the respective
%line strengths. In both cases the solid line indicates the best fit.}
%\label{fig:6}
%\end{figure}

%\begin{figure}
%\centering
%\includegraphics[width=8.5cm,height=6.5cm]{v1_y2isochrone_gsc.ps}
%\caption{Theoretical isochrones from \citet{Demarque2004ApJS}
%for different ages;  the black line corresponds to
%an age of 12 Gyr, whereas the violet and red lines correspond to 11 Gyr and 13 Gyr, respectively.
%We have assumed [Fe/H] = $3.5$ and [$\alpha$/Fe] = 0.0. The log $\textit{g}$
%values are taken where the temperature lines intersects the isochrones. 
%For the adopted $T_{eff}$ (5620~K), indicated by the vertical green line, 
%we find two possible log $\textit{g}$ 
%values (log $\textit{g}$ = 4.7,3.4). }
%\label{fig:3}
%\end{figure}

\subsection{ Abundance analysis}

We have used the ATLAS9 NEWODF \citep{castelli-kurucz2004astroph} model
atmospheres without convective overshooting. The main abundance analysis
results come from the APO spectrum, since it covers a broader wavelength
region. In the overlapping regions of both the APO and HET spectra, a
final abundance value is derived using a weighted average of abundances
from both spectra. For the H$_{\beta}$ and Mg fitting, the HET spectrum
is primarily used because these lines fall at the edge of the echelle
orders, and in such cases continuum normalization was better
accomplished from the HET spectrum. We considered only lines with
equivalent widths less than 100 m\AA\ for the abundance analysis, since
they are on the linear part of the curve of growth, and are relatively
insensitive to the choice of microturbulence.

We have adopted the solar abundance from \citet{Grevesse1998}; solar
isotopic fractions were used for all the elements. We used the CH
linelist compiled by Masseron (private communication) and CN molecular
line list compiled by Plez et al. (2005), and the NH and C$_{2}$
molecular-line lists from the Kurucz database. The atomic data for the
other lines have been listed in the appendix along with the respective
references used. The line lists used are the same as in
\citet{Cui2013AA}. We employed version 12 of the turbospectrum
code for spectrum synthesis and abundance analysis
\citep{AlvarezBplez1998AA,Plez2012ascl}.

\begin {table}
\centering
\caption{The adopted stellar parameters for SDSS~J1343+4844 }
\begin{tabular}{c c c c}
\hline
$T_{eff}$   & log\textit {g}    & $\xi$ & [Fe/H]    \\
(K)		 & (cgs)    &	(km/s)  & dex	   \\
\hline\hline
5620		&  3.44	   & 1.45     & $-3.42$   \\	
\hline
\end{tabular}
\label{tab:5}
\end{table}  

{\footnotesize
\begin{table}
\centering
\caption{Elemental abundance determinations for SDSS~J1343+4844}
\begin{tabular}{ccrcrrc}
\hline\hline
Species & $N_{lines}$&   A(X)   &   Solar &  [X/H]   & [X/Fe]$^{\#}$	    & $\sigma^{*}$\\
\hline

CH 	& \dots	     &  5.52 	&  8.52   & $-$3.0     & $+$0.42	    & synth \\
Na~I 	&  2         &	2.32    &  6.33   & $-$4.01    & $-$0.59$^{\mathsection}$  & 0.01   \\	%without NLte 2.62
Mg~I	&  3	     &  4.31    &  7.58   & $-$3.27    & $+$0.15	    & 0.02	 \\
Al~I    &  2         &	3.30    &  6.47   & $-$3.15    & $+$0.25$^{\mathsection}$   & 0.07	\\	%without NLTE 2.65
Si~I	&  1         &	4.11    &  7.55   & $-$3.44    & $-$0.02 &	\dots\\  
S ~I	& \dots	     & $<$5.64	&  7.33	  & $< -$1.69  & $< +$1.73$^\dagger$ &  synth  \\
Ca~I  	&  4	     &  3.13    &  6.36   & $-$3.23    & $+$0.19	    &	0.16   \\
Sc~II	&  1	     & $-0.08$  &  3.17   & $-$3.25    & $+$0.17	    &  \dots \\
Ti~II  	&  11	     &  1.98    &  5.02   & $-$3.04    & $+$0.38 	    & 0.25   \\
Cr~I	&  2	     &  2.28    &  5.67   & $-$3.39    & $+$0.03	    & 0.04    \\ 
Mn~I	&  2	     &  1.75    &  5.39   & $-$3.64    & $-$0.22	    &	0.10\\
Fe~I 	&  60	     &  4.08    &  7.50   & $-$3.42    & 0.00	            & 0.19	\\
Fe~II   &  4         &  4.37    &  7.50   & $-$3.13    & 0.29               & 0.08  \\
Ni~I	&  3	     &  3.15    &  6.25   & $-$3.10    & $+$0.32	    & 0.10 \\ 
Sr~II	&  2	     & $-0.92$  &  2.97   & $-$3.89    & $-$0.47	    &   0.12       \\
Ba~II	& \dots	     &$< -1.83$ &  2.13   & $<-$3.96   & $< -$0.54$^\dagger$& synth	\\
\hline
\end{tabular}

$^\mathsection$ Represents the values after applying the NLTE corrections.\\
$^\dagger$ Denotes the abundance values from the synthesis.~~~~~~~~~~~~~~~~ \\ 
$^*$ The errors quoted are random errors.~~~~~~~~~~~~~~~~~~~~~~~~~~~~~~~~~\\
$^{\#}$ [X/Fe] is calculated using the Fe I abundance.~~~~~~~~~~~~~~~~~~~~~~~ \\
\end{table}
}

\section{Abundances}

\subsection{Carbon, Nitrogen, and Oxygen}

There is only a weak CH G-band present in our spectra. The carbon
abundance was derived by iteratively fitting the bandhead region with
synthetic spectra, and adopting the value that yields the best matched
the observed spectra. The APO spectrum is very noisy in the G-band
region, and the HET spectrum did not cover this wavelength. So, we have
used the low-resolution SDSS and HCT spectra to derive the carbon
abundance from the CH G-band region. The fit is not particularly good
due to the weakness of the band and contamination due to the H$\gamma$
wing, so our reported value should be considered provisional. However,
it is also obvious that \gscc\ is not enhanced in carbon, which would
have resulted in a much larger CH G-band for the S/N of the observed
spectra. The signal to noise ratio at 338 nm wavelength region is very
poor to confirm an enhancement in nitrogen abundance. Oxygen lines at
777 nm is too weak to be detected. Hence we could not derive meaningful
abundances.

\subsection{ The $\alpha$-elements}

Magnesium lines at 3838\,{\AA} and 5172\,{\AA} are detected in the
spectra. Only two lines of the triplet at 5172 \AA\, and one clean line
in the 3838 \AA\ region were used to determine the abundance. The
derived [Mg/Fe] ratio is ([Mg/Fe] = +0.15); it is not enhanced as often
found for halo stars. The silicon line at 3905\,{\AA} is used for the
abundance estimate; the result is also close to solar, [Si/Fe] =
$-0.02$. The titanium abundance is found to be [Ti/Fe] = +0.38. We could not
detect the S~I lines at 9212\,{\AA} and 9237\,{\AA}, thus only derived
an upper limit for sulphur. The Ca abundance was derived from the three
prominent Ca~I lines at 4226.73\, {\AA}, 4302.53\,{\AA} and 4434.96\,
{\AA}; they indicate a slightly enhanced [Ca/Fe] ratio ($+0.19$). 

Overall, the $\alpha$-element enhancement for SDSS~J1343+4844
([$\alpha$/Fe] $\sim$+0.1, taking only the Mg and Si abundances into
account) is lower than for typical halo stars in the Galaxy
([$\alpha$/Fe] $\sim +0.4$). Non-LTE effects on the Mg and Ca abundances
for EMP stars were investigated by \citet{Mashonkina2007AA},
\citet{Andrievsky2010AA}, and \citet{Spite2012AA}. The non-LTE corrections 
for Mg are +0.1 to +0.3 dex. Hence, the Mg abundance ratios could be
systematically higher than those derived by our LTE analysis. However,
most of the Mg abundances results that are available for EMP stars are
also based on an LTE analysis, and exhibit higher [Mg/Fe] abundance
ratios relative to SDSS~J1343+4844. The Ca~I 4226\,{\AA} line is
affected by NLTE effects, compared to other subordinate lines. These
lines are very weak in our spectra, however they show consistent values
with the abundances derived from Ca~I 4226\, {\AA}.

\subsection{The odd-Z elements}

The sodium abundance is derived from the Na $D_1$ and $D_2$ resonance
lines at 5890\,{\AA} and 5896\,{\AA}, and the abundance of aluminium is
derived using the resonance doublet at 3944\,{\AA} and 3961.5\,{\AA}.
These resonance lines are very sensitive to NLTE effects. The NLTE
corrections to the abundance are $-0.30$ dex for Na
\citep{Baumueller1998} and $+0.65$ dex for Al \citep{Baumueller1997}.

After the NLTE corrections are applied, our derived abundances for this
star are in good agreement with the values found for other EMP stars
\citep{Cayrel2004,Lai2008ApJ}. 

\subsection{The iron-peak elements}

Iron abundances are derived from 68 Fe~I and 4 Fe~II lines. The
abundance derived from the Fe~I and Fe~II lines show a difference of
$\sim$ 0.3 dex, which may arise from the small number of weak Fe~II lines. The
difference between Fe~I and Fe~II is in agreement with the NLTE effects explored by
\citet{Asplund2005ARA&A}. We detected Mn, Cr, and Ni among the iron-peak
elements. The abundance of Mn was derived from the resonance Mn triplet
at 4030\,{\AA}. The spectral line at 4034.4\,{\AA} was not used in the
abundance analysis, since it is affected by a bad pixel.
The abundance of scandium is based on one line at 4246.8\,{\AA},
and is found to be similar to other EMP stars.  The observed
abundances of Cr and Mn are also similar to other EMP stars. The [Cr/Mn] ratio
at the metallicity of SDSS J1343+4844 is in agreement with the [Cr/Mn]
ratio obtained for EMP stars by \citet{Cayrel2004}. The abundance ratio
derived for Ni, using two lines (3783.5\,{\AA} and 3858.29\, {\AA}) is
[Ni/Fe] = $+$0.32.
%, which is also similar to the sample from
%\citet{Cayrel2004}.

\begin{figure}
\centering
\includegraphics[width=8.5cm, height=6.5cm]{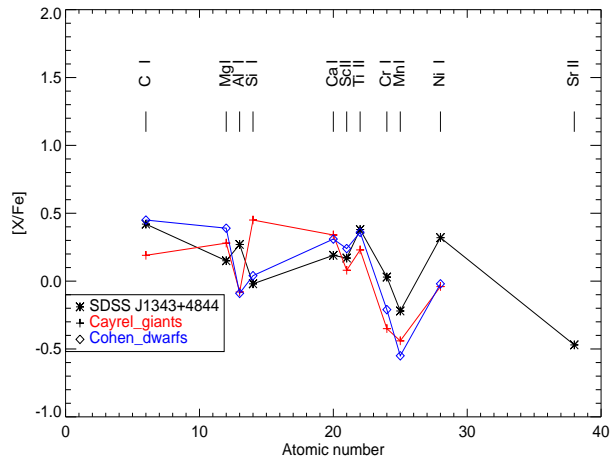}
\caption{Elemental abundance for SDSS~J1343+4844, compared with samples available from \citet{Cayrel2004}
and \citet{Cohen2004ApJ}. The red crosses indicate the mean elemental abundances from the Cayrel
sample and the open diamonds represent the mean abundance values of the Cohen sample (EMP stars).
The black asterisks represent the abundances of \gscc. }
\label{fig:compare_samp}
\end{figure}

Figure \ref{fig:compare_samp} shows a comparison of our derived
abundances for SDSS~J1343+4844 compared to the mean abundances of EMP
giants and dwarfs. The iron-peak elements agree with the mean EMP
abundances of these elements.

\subsection{The neutron-capture elements}

The neutron-capture elements in SDSS~J1343+4844 are under-abundant
relative to the solar values. 

Resonance lines of strontium Sr~II 4077\, {\AA} and 4215\,{\AA} are
detected; both the lines exhibit under-abundances with respect to the
solar ratio ([Sr/Fe] = $-0.47$). We did not detect barium, thus we
estimated an upper limit for the abundance of Ba in \gscc\ by fitting
the Ba~II resonance lines at 4554\,{\AA} and 4934\,{\AA}, incorporating
the hyperfine splitting by \citet{McWilliam1995AJII} and assuming a
solar isotopic composition. The upper limit derived for Ba is [Ba/Fe] $<
-0.54$, also quite low.

%\begin{figure}
%\centering
%\includegraphics[width=8.5cm, height=6.5cm]{test_carb_review_report_long_withsdss.ps}
%\caption{The CH G-band at 4300\,{\AA} fitted for various values of carbon abundance.
%The red line indicates the best fit, where the absolute carbon abundance
%used is log $\epsilon$ = 5.52. We used a step size of 0.3 dex from the
%best-fit value, represented by the blue and green colours. The
%observed HCT spectra are plotted in black and magenta colours.
%The SDSS spectrum is also plotted but in violet color.
%  The G-band
%feature is not particularly well-fit, but it is clearly weak.}
%\textbf{The 
%4271 feature in the synthetic spectra is the Fe I line. It is poorly fitted 
%because the observed spectra is of low resolution.}}
%\label{carb_G}
%\end{figure}

\begin{figure}
\centering
\includegraphics[width=8.5cm, height=6.5cm]{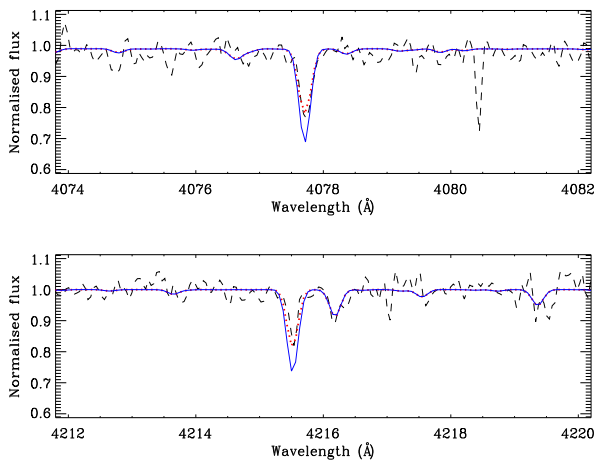}
\caption{The Sr~II lines at 4077\,{\AA} and 4216\,{\AA} are shown.
The APO spectrum is plotted as a black dashed line,
The abundance value derived from the equivalent-width analysis is used for the synthetic spectrum
 and plotted as a red dotted line. The blue solid line indicates the synthesized spectrum for
[Sr/Fe] = 0.
}
\end{figure}

\section{Discussion and conclusions}

\subsection{CNO abundances in unevolved  stars with [Fe/H] $< -3.0$}

Metal-poor stars with enhanced/peculiar abundances in certain key
elements help to identify the nuclear processes and astrophysical sites
of their production. However, it is also interesting to study objects
that do not exhibit such enhancements, since they may represent the mean
ISM abundances at those epochs. This will help to understand the
frequency of the nuclear sites that produce the peculiar abundance
enhancements. SDSS~J1343+4844 does not exhibit enhancement of  C and
possibly N and O abundances, indicating that there was no internal CN processing,
which is consistent with its evolutionary stage as a subgiant.
\citet{Spite2006A&A} and \citet{Bonifacio2009A&A} show that the mean
[C/Fe] abundance ratios in unmixed giants and dwarfs are about [C/Fe] =
+0.19 and [C/Fe] = +0.45, respectively. The SAGA compilation
\citep{Suda2008PASJ} and Figure~2 of \citet{Bonifacio2009A&A} also show
that the carbon abundance ratios in the carbon-normal EMP stars are
slightly enhanced with respect to the solar ratio. SDSS~J1343+4844 has
[C/Fe] = $+$0.42, based on the low-resolution spectrum, consistent with
these EMP stars.
 
\subsection{Stars with low $\alpha$-element abundances at [Fe/H] $< -3.0$}

In spite of having similar elemental abundances compared with the mean
abundances of EMP stars, SDSS~J1343+4844 belongs to the rare class of
carbon-normal EMP stars that exhibit low $\alpha$-element abundances.
It is thus interesting explore the origin of such stars.

\begin{figure}
\centering
\includegraphics[width=8.0cm, height=6.0cm]{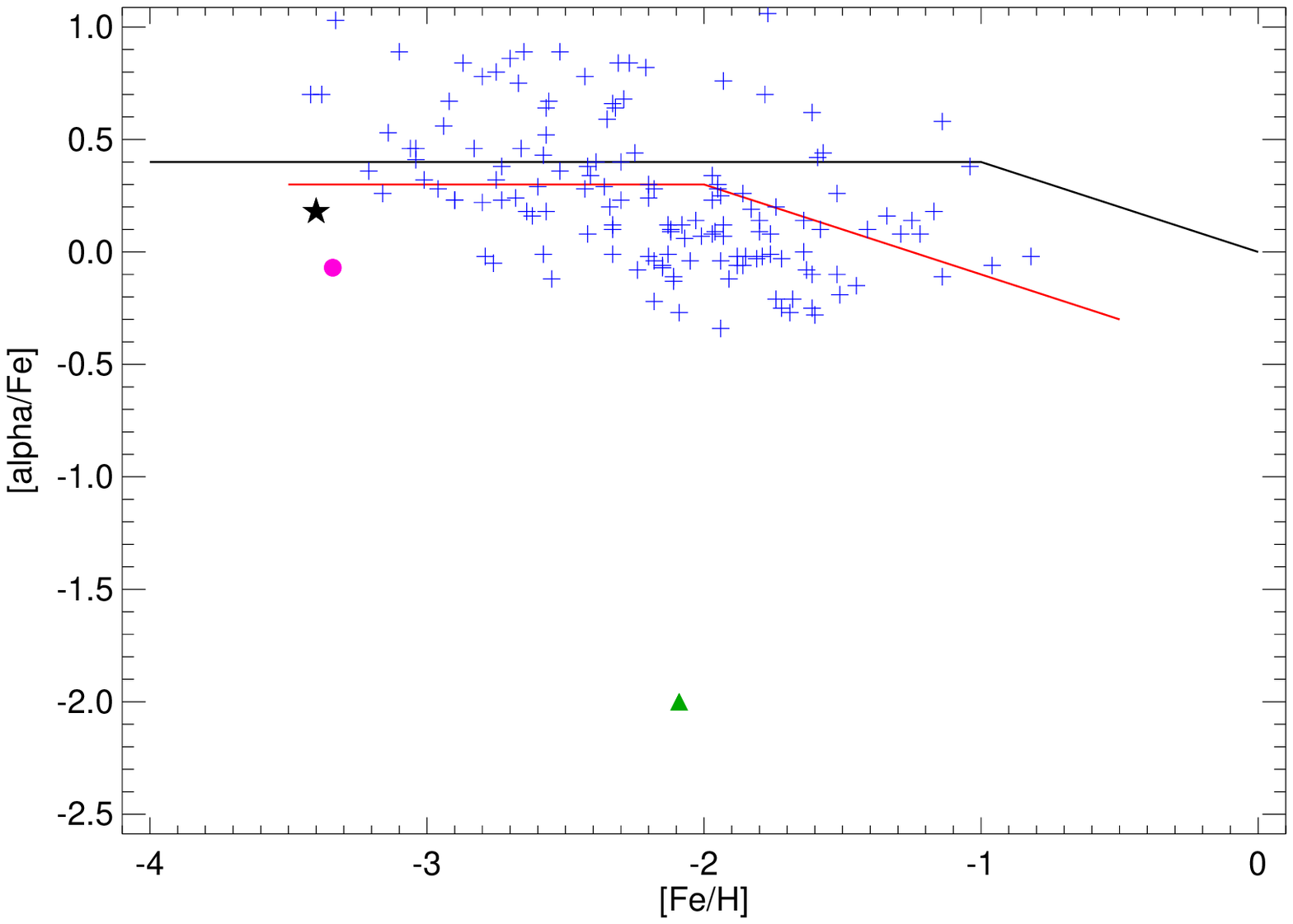}
\caption{$\alpha$-elemental abundance for \gscc\ (black star), compared with samples 
available from \citet{Ivans2003ApJ} (green filled triangle),\citet{vargas_2013ApJ}
(the blue crosses),\citet{Caffau2013AA} (magenta filled circle). The black line
is the schematic representation of the mean $\alpha$ abundance of the
Milky Way halo, thick disk, and thin disk. The red line is a schematic
representation of the mean $\alpha$ abundance of dwarf spheroidal
galaxies. }
\label{fig:abund1}
\end{figure}

Figure \ref{fig:abund1} shows the abundances of elements from the
\citet{vargas_2013ApJ} sample. The $\alpha$-poor stars from
\citet{Ivans2003ApJ} and \citet{Caffau2013AA} are also shown.  
\gscc\ falls in the region close to the mean $\alpha$ abundances of dwarf galaxies.
However the halo $\alpha$-element abudances for stars with  [Fe/H] $< -3.0$
exhibit a large scatter.

% $\alpha$ elements are ejected to ISM mainly due to SNe II explosions. 
%Majority of EMP samples in the halo as well as some dwarf galaxies share a common pattern
%for these $\alpha$ elements. Neverthless, there have been studies which
% reported EMP $\alpha$-poor population in the halo and the nearby dSphs.
%%The existence of such stars points to either a different astrophysical site and/or inefficient mixing mechanism. 
%shows a different formaton history compared to the normal population.
%\citet{Ivans2003ApJ}}
%The $\alpha$-poor nature of EMP stars listed in the figure ~\ref{abund1} could not be
%explained by the SNIa scenario as their occurance is much later than the 
%EMP star formed in the Galaxy.
 
Low $\alpha$-element abundances are expected for a low star-formation
rate, similar to that found for many dwarf spheroidal galaxies. If this
star belongs to a population that is representative of a low
star-formation rate, inhomogeniously-mixed system, then SDSS~J1343+4844
could be younger than other EMP stars, which are $\alpha$-element
enhanced at the same metallicity.

\subsection{Neutron-capture elements for stars with [Fe/H] $< -3.0$ }

SDSS~J1343+4844 exhibits similar abundance patterns as those of EMP
dwarfs and giants for the Fe-peak elements and n-capture elements as
well. Strontium abundance ratios for EMP dwarfs and giants are between
[Sr/Fe] = $-$0.6 and [Sr/Fe] = $-$0.1 at metallicities of [Fe/H] =
$-3.0$ \citep{Andrievsky2011AA}, which is again similar to
SDSS~J1343+4844. The upper limit for the Ba abundance of our star ([Ba/Fe]
$< -0.54$) lies well within the average Ba abundances of EMP stars
having normal carbon abundance at the same metallicity
(\citealt{Cohen2013ApJ}; [Ba/Fe] $\sim -0.6$). 

During the early epochs of chemical evolution, asymptotic giant-branch
stars would not have had time to contribute to the ISM
\citep{Kobayashi2014ApJ}. They contribute significantly only after z =
1.8 (roughly [Fe/H] $> -$2.4). Hence, the n-capture elements seen in
SDSS~J1343+4844 have likely had contributions from massive stars. Recent
observations of large opacity due to heavy elements in kilonovae
\citet{tanvir2013Nature} has brought renewed interest in NS-NS mergers
as a promising candidate for $r$-process-element production. NS-NS mergers were
proposed long ago \citep{lattimer_schramm1976ApJ, lattimer1977ApJ,
Meyer1989ApJ}, and also in recent studies \citep{freiburghaus1999ApJ,
Goriely2011ApJ, rosswog2014MNRAS, wanajo2014ApJ,vangioni2016MNRAS}, as
promising sites for production of the $r$-process elements.
Observationally-motivated work on this subject includes
\citet{honda2004ApJ}, \citet{Francois2007AA}, and
\citet{sneden2008ARAA}. These spectroscopic studies find a large scatter
in [Eu/Fe] ratios (at least two orders of magnitude) for stars with
[Fe/H] $\le$ -2.5. They also show that the lighter and heavier
$r$-process elements correlate very well, indicating possible
connection, however their ratios at different metallicites suggest two
possible evolutionary behaviours below [Fe/H] $< -2.5$. An additional
mechanism, such as the LEPP (light element primary process), might be
operating at metallicities [Fe/H] $< -3.0$ (Montes et al. 2007; Francois
et al. 2007)

Most chemical evolution models are unable to explain all the
observations. Recent work by \citet{Tsujimoto_Shigeyama2014ApJ} and
\citet{Ishimaru_Wanajo_Prantzos_2015ApJ} address these issues by
assuming a chemical evolution of sub-halos with different masses and
star-formation rates (SFR) contributing to the Galactic halo rather than
chemical evolution within a single halo.
\citet{prantzos2006astro.ph.12633P} showed that if the sub-halos evolved
at different rates, there would be no unique relation between age and
metallicity; in that case there is possibility of an "early" appearance
of $r$-process elements and their large dispersion can be explained,
even if the main source of those elements is NSMs.

\begin{figure}
\centering
\includegraphics[width=8.0cm, height=6.0cm]{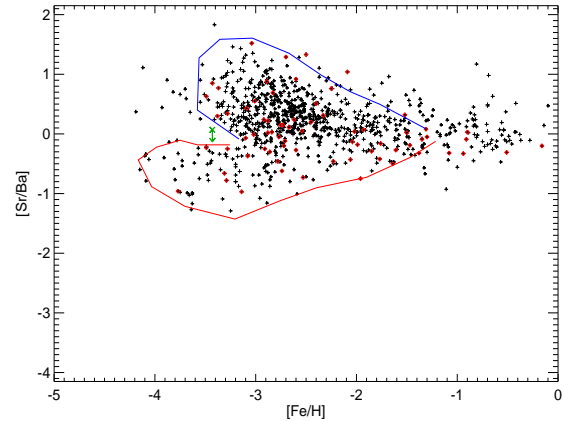}
\caption{ The evolution of lighter versus heavier $r$-process 
elemets are shown. The blue and red lines mark 
the contributions from massive and low-mass sub-halos that 
were accreted into the Galaxy, similar to Tsujimoto \& Shigeyama (2015).
The high ratios of [Sr/Ba] are due to massive sub-halos, while low [Sr/Ba]
values are from low-mass sub-halos. These are non-CEMP stars, 
hence $s$-process contribution from stars with [Fe/H] $\le -2.0$
is not expected. The data is from the SAGA database
\citep{Suda2008PASJ}. \gscc\ is marked as a green 'X'. }
\label{fig:r-process_in alpha} 
\end{figure}

\begin{figure}
\centering
\includegraphics[width=8.5cm, height=6.5cm]{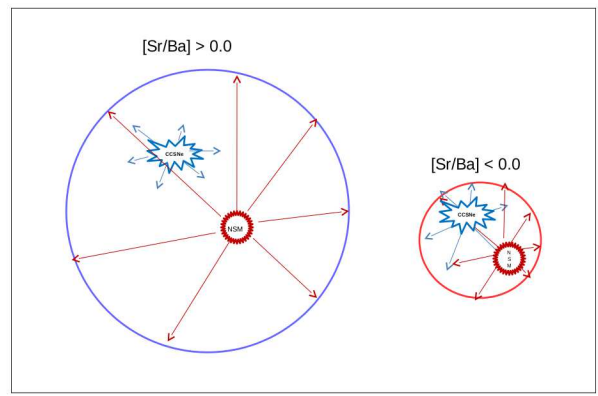}
\caption{A schematic representation of the 
mechanism proposed by \citet{Tsujimoto_Shigeyama2014ApJ}. The wind from
NS-MS mergers is shown in red and the wind from CCSNe are shown in blue.
The wind from NS-NS mergers span the entire sub-halo, whereas the wind
from CCSNe blast waves does not cover the entire massive sub-halo
(10$^{9}$M$_\odot$), but pollutes a larger filling factor for a low-mass
halo (10$^{6}$-10$^{7}$M$_\odot$), producing different dilution and
hence different ratios of [Sr/Ba].}
\label{fig:11_new}
\end{figure}

Figure \ref{fig:r-process_in alpha} is motivated by Figure 2 in 
\citet{Tsujimoto_Shigeyama2014ApJ}.
\citet{Tsujimoto_Shigeyama2014ApJ} successfully explained 
the large scatter in [Eu/Fe] at low metallicities and the 
observed correlation between [light $r$-process/Eu] and [Eu/Fe].
According to these authors, CCSNe produce lighter $r$-process elements 
and the heavier $r$-process is produced by NS-NS mergers. 
The nucleosynthesis products of these two sources are mixed
to  different amounts within the sub-halos due to the differences
in their wind speeds and a variations of sub-halo masses. 
A schematic diagram, shown in Figure \ref{fig:11_new}, illustrates this.
%The fast winds from NS-NS mergers will sweep through the
%entire halo, producing different dilution for different 
%masses of the sub-halos, whereas the slow winds from CCSNe
%will have similar level of mixing of ISMs of different 
%sub-halo masses, producing different ratios of
%lighter and heavier $r$-process ratios.  
The high ratios of [light $r$-process/heavier $r$-process] is argued to
be produced by massive sub-halos and
a lower ratio is produced in low mass sub-halos. 

Here we use barium to represent heavier $r$-process elements, and
strontium to represent lighter $r$-process elements. The data shown in
Figure \ref{fig:r-process_in alpha} do not include carbon-enhanced stars, hence $s$-process
contribution is not expected for stars with [Fe/H] $< -2.0$. The two
regimes shown by \citet{Tsujimoto_Shigeyama2014ApJ} are clearly seen in
this figure as well. The region enclosed by the blue curve is expected
to be occupied by stars that are formed within massive sub-halos
(10$^{9}$M$_\odot$) and the region enclosed by the red curve corresponds
to stars that are formed within low-mass sub-halos
(10$^{6}$-10$^{7}$M$_\odot$). \gscc\ clearly falls in the middle of
these distribution, indicating that the star is formed from a mixed ISM
without peculiar composition. Figure \ref{fig:r-process_in alpha}
 also shows the location of
$\alpha$-poor stars with [$\alpha$/Fe] $<$ 0.1 (red dots). We find that
they do not occupy any preferred location; rather, they are uniformly
spread, in-spite of the two regions shown in the plot representing
different mass ranges of the sub-halos. This result indicates that the
SFRs for sub-halos of different masses may be similar during these
epochs. Normally, $\alpha$-poor stars are expected to have formed in
low-mass sub-halos with low SFRs. A different SFR for different sub-halo
masses will also affect the [Sr/Ba] ratio, which does not match well
with the observations. Apart from the sub-halo masses and SFR, the
timescale of these sub-halos merging into the Galaxy is also essential
for understanding the final abundance ratios.

\subsection{Lithium depletion and binarity}

The lithium line at 6707\,{\AA} is not detected in SDSS~J1343+4844. The
derived upper limit agrees with the declining lithium abundance trend
for stars with [Fe/H] $< -3.0$ \citep{sbordone2010AA}. There are two
possibilities for the depletion of lithium: binary mass transfer and
binary-induced mixing, or mixing due to the evolutionary status of the
object as a subgiant. Detection of a radial-velocity variation with no
peculiar chemical composition support the idea of binary-induced mixing.
Most short-period binaries are found to have preserved their lithium,
possibly due to tidal locking. However long-period binaries with periods
of few hundred days seem to have depleted lithium abundances
\citep{Ryan1995ApJ}, which could be the case for SDSS~J1343+4844.

\subsection{Conclusions}

We have derived LTE abundances for SDSS~J1343+4844, a bright EMP star
identified during the MARVELS pre-survey. The abundance pattern for most
elements in this star are similar to other carbon-normal EMP stars,
except for the low $\alpha$-element abundance and the apparent depletion
of lithium. The depletion of lithium could be understood in terms of
binary-induced mixing or additional convection, as the object is a
subgiant. Low $\alpha$-element abundance, along with the agreement of
the observed ratios of the carbon abundances and heavy elements with the
mean EMP value is not so easy to understand. We also show that the
$\alpha$-poor halo stars exhibit a wide range of lighter and heavier
$r$-process ratios. An ISM with contributions from Pop III
intermediate-mass stars, along with later Pop II contributions with a
low SFR, may explain the abundance patterns seen in \gscc\.%
SDSS~J1343+4844.

Predictions of several key elemental
abundance ratios, especially C,N, $\alpha$ and light/heavy $r$-process
elements from a chemical evolution model incorporating hierarchical merging of
sub-halos of different masses and SFR may help to understand these
observations.

%We have shown that the elemental abundances of \gscc\ may reflect
%the ISM mean abundances along with other $\alpha$-poor nature.

%We have also shown that the $r$-process ratios
%along with $\alpha$-element abundances  for different Milky Way satellites might be important 
%to understand these observations, within the hierarchical galaxy formation scenario.
%Though we propose the star
%could have formed in a different system where the star formation rate is low and the ISM
%remained unmixed until the onset of SNe Ia,
%and accreted to our galaxy through merging scenario, We need more samples to further
% constrain the argument and need better models for a detailed explanation to the physical
%processes existed in the early universe.}

\subsection{acknowledgements}

TCB acknowledges partial support for this work from grant 
PHY 14-30152; Physics Frontier Center/JINA Center
for the Evolution of the Elements (JINA-CEE), awarded by the US National
Science Foundation. The authors thank Thomas Masseron and Bertrand Plez for sharing the
updated CH linelist.  Many thanks to the referee, Chris Sneden, for his
valuable inputs and willingness to discuss the paper before 
resubmission of the manuscript, which improved the paper considerably.
SDSS-III MARVELS pre-survey data is used in this work.
Funding for SDSS-III has been provided by the Alfred P. Sloan Foundation, 
the Participating Institutions, the National Science Foundation, and the U.S.
Department of Energy Office of Science. The SDSS-III web site is http://www.sdss3.org/.

SDSS-III is managed by the Astrophysical Research Consortium for the
Participating Institutions of the SDSS-III Collaboration including the
University of Arizona, the Brazilian Participation Group, Brookhaven
National Laboratory, Carnegie Mellon University, University of Florida,
the French Participation Group, the German Participation Group, Harvard
University, the Instituto de Astrofisica de Canarias, the Michigan
State/Notre Dame/JINA Participation Group, Johns Hopkins University,
Lawrence Berkeley National Laboratory, Max Planck Institute for
Astrophysics, Max Planck Institute for Extraterrestrial Physics, New
Mexico State University, New York University, Ohio State University,
Pennsylvania State University, University of Portsmouth, Princeton
University, the Spanish Participation Group, University of Tokyo,
University of Utah, Vanderbilt University, University of Virginia,
University of Washington, and Yale University.

\bibliographystyle{mn2e}
\bibliography{papergsc_ref}

\begin{thebibliography}{80}
\expandafter\ifx\csname natexlab\endcsname\relax\def\natexlab#1{#1}\fi

\bibitem[{{Alonso}, {Arribas} \& {Martinez-Roger}(1996){Alonso}, {Arribas}, \&
  {Martinez-Roger}}]{Alonso1996}
{Alonso} A., {Arribas} S., {Martinez-Roger} C., 1996, \aap, 313, 873

\bibitem[{{Alonso}, {Arribas} \& {Mart{\'{\i}}nez-Roger}(1999){Alonso},
  {Arribas}, \& {Mart{\'{\i}}nez-Roger}}]{Alonso1999}
{Alonso} A., {Arribas} S., {Mart{\'{\i}}nez-Roger} C., 1999, \aaps, 139, 335

\bibitem[{{Alvarez} \& {Plez}(1998)}]{AlvarezBplez1998AA}
{Alvarez} R., {Plez} B., 1998, \aap, 330, 1109

\bibitem[{{Anderson}, {Zilitis} \& {Sorokina}(1967){Anderson}, {Zilitis}, \&
  {Sorokina}}]{AZS}
{Anderson} E.~M., {Zilitis} V.~A., {Sorokina} E.~S., 1967, Optics and
  Spectroscopy, 23, 102, (AZS)

\bibitem[{{Andrievsky} {et~al}\mbox{.}(2011){Andrievsky}, {Spite}, {Korotin},
  {Fran{\c c}ois}, {Spite}, {Bonifacio}, {Cayrel}, \&
  {Hill}}]{Andrievsky2011AA}
{Andrievsky} S.~M., {Spite} F., {Korotin} S.~A., {Fran{\c c}ois} P., {Spite}
  M., {Bonifacio} P., {Cayrel} R., {Hill} V., 2011, \aap, 530, A105

\bibitem[{{Andrievsky} {et~al}\mbox{.}(2010){Andrievsky}, {Spite}, {Korotin},
  {Spite}, {Bonifacio}, {Cayrel}, {Fran{\c c}ois}, \&
  {Hill}}]{Andrievsky2010AA}
{Andrievsky} S.~M., {Spite} M., {Korotin} S.~A., {Spite} F., {Bonifacio} P.,
  {Cayrel} R., {Fran{\c c}ois} P., {Hill} V., 2010, \aap, 509, A88

\bibitem[{{Aoki} {et~al}\mbox{.}(2009){Aoki}, {Arimoto}, {Sadakane}, {Tolstoy},
  {Battaglia}, {Jablonka}, {Shetrone}, {Letarte}, {Irwin}, {Hill}, {Francois},
  {Venn}, {Primas}, {Helmi}, {Kaufer}, {Tafelmeyer}, {Szeifert}, \&
  {Babusiaux}}]{Aoki2009AA}
{Aoki} W. {et~al.}, 2009, \aap, 502, 569

\bibitem[{{Asplund}(2005)}]{Asplund2005ARA&A}
{Asplund} M., 2005, \araa, 43, 481

\bibitem[{{Barklem} {et~al}\mbox{.}(2005){Barklem}, {Christlieb}, {Beers},
  {Hill}, {Bessell}, {Holmberg}, {Marsteller}, {Rossi}, {Zickgraf}, \&
  {Reimers}}]{Barklem2005AA}
{Barklem} P.~S. {et~al.}, 2005, \aap, 439, 129

\bibitem[{{Baumueller}, {Butler} \& {Gehren}(1998){Baumueller}, {Butler}, \&
  {Gehren}}]{Baumueller1998}
{Baumueller} D., {Butler} K., {Gehren} T., 1998, \aap, 338, 637

\bibitem[{{Baumueller} \& {Gehren}(1997)}]{Baumueller1997}
{Baumueller} D., {Gehren} T., 1997, \aap, 325, 1088

\bibitem[{{Beers} \& {Christlieb}(2005)}]{beers-christlieb2005ARA&A}
{Beers} T.~C., {Christlieb} N., 2005, \araa, 43, 531

\bibitem[{{Beers} {et~al}\mbox{.}(2014){Beers}, {Norris}, {Placco}, {Lee},
  {Rossi}, {Carollo}, \& {Masseron}}]{tcbeers2014ApJ}
{Beers} T.~C., {Norris} J.~E., {Placco} V.~M., {Lee} Y.~S., {Rossi} S.,
  {Carollo} D., {Masseron} T., 2014, \apj, 794, 58

\bibitem[{{Bonifacio} {et~al}\mbox{.}(2009){Bonifacio}, {Spite}, {Cayrel},
  {Hill}, {Spite}, {Fran{\c c}ois}, {Plez}, {Ludwig}, {Caffau}, {Molaro},
  {Depagne}, {Andersen}, {Barbuy}, {Beers}, {Nordstr{\"o}m}, \&
  {Primas}}]{Bonifacio2009A&A}
{Bonifacio} P. {et~al.}, 2009, \aap, 501, 519

\bibitem[{{Bromm}, {Yoshida} \& {Hernquist}(2003){Bromm}, {Yoshida}, \&
  {Hernquist}}]{Bromm2003ApJ}
{Bromm} V., {Yoshida} N., {Hernquist} L., 2003, \apjl, 596, L135

\bibitem[{{Caffau} {et~al}\mbox{.}(2013){Caffau}, {Bonifacio}, {Fran{\c c}ois},
  {Sbordone}, {Spite}, {Monaco}, {Plez}, {Spite}, {Zaggia}, {Ludwig}, {Cayrel},
  {Molaro}, {Randich}, {Hammer}, \& {Hill}}]{Caffau2013AA}
{Caffau} E. {et~al.}, 2013, \aap, 560, A15

\bibitem[{{Carney} {et~al}\mbox{.}(1997){Carney}, {Wright}, {Sneden}, {Laird},
  {Aguilar}, \& {Latham}}]{Carney1997AJ}
{Carney} B.~W., {Wright} J.~S., {Sneden} C., {Laird} J.~B., {Aguilar} L.~A.,
  {Latham} D.~W., 1997, \aj, 114, 363

\bibitem[{{Castelli} \& {Kurucz}(2004)}]{castelli-kurucz2004astroph}
{Castelli} F., {Kurucz} R.~L., 2004, ArXiv Astrophysics e-prints

\bibitem[{{Cayrel} {et~al}\mbox{.}(2004){Cayrel}, {Depagne}, {Spite}, {Hill},
  {Spite}, {Fran{\c c}ois}, {Plez}, {Beers}, {Primas}, {Andersen}, {Barbuy},
  {Bonifacio}, {Molaro}, \& {Nordstr{\"o}m}}]{Cayrel2004}
{Cayrel} R. {et~al.}, 2004, \aap, 416, 1117

\bibitem[{{Cohen} {et~al}\mbox{.}(2004){Cohen}, {Christlieb}, {McWilliam},
  {Shectman}, {Thompson}, {Wasserburg}, {Ivans}, {Dehn}, {Karlsson}, \&
  {Melendez}}]{Cohen2004ApJ}
{Cohen} J.~G. {et~al.}, 2004, \apj, 612, 1107

\bibitem[{{Cohen} {et~al}\mbox{.}(2013){Cohen}, {Christlieb}, {Thompson},
  {McWilliam}, {Shectman}, {Reimers}, {Wisotzki}, \& {Kirby}}]{Cohen2013ApJ}
{Cohen} J.~G., {Christlieb} N., {Thompson} I., {McWilliam} A., {Shectman} S.,
  {Reimers} D., {Wisotzki} L., {Kirby} E., 2013, \apj, 778, 56

\bibitem[{{Cui}, {Sivarani} \& {Christlieb}(2013){Cui}, {Sivarani}, \&
  {Christlieb}}]{Cui2013AA}
{Cui} W.~Y., {Sivarani} T., {Christlieb} N., 2013, \aap, 558, A36

\bibitem[{{Cutri} {et~al}\mbox{.}(2003){Cutri}, {Skrutskie}, {van Dyk},
  {Beichman}, {Carpenter}, {Chester}, {Cambresy}, {Evans}, {Fowler}, {Gizis},
  {Howard}, {Huchra}, {Jarrett}, {Kopan}, {Kirkpatrick}, {Light}, {Marsh},
  {McCallon}, {Schneider}, {Stiening}, {Sykes}, {Weinberg}, {Wheaton},
  {Wheelock}, \& {Zacarias}}]{Cutri2003}
{Cutri} R.~M. {et~al.}, 2003, VizieR Online Data Catalog, 2246, 0

\bibitem[{{Demarque} {et~al}\mbox{.}(2004){Demarque}, {Woo}, {Kim}, \&
  {Yi}}]{Demarque2004ApJS}
{Demarque} P., {Woo} J.-H., {Kim} Y.-C., {Yi} S.~K., 2004, \apjs, 155, 667

\bibitem[{{Fran{\c c}ois} {et~al}\mbox{.}(2007){Fran{\c c}ois}, {Depagne},
  {Hill}, {Spite}, {Spite}, {Plez}, {Beers}, {Andersen}, {James}, {Barbuy},
  {Cayrel}, {Bonifacio}, {Molaro}, {Nordstr{\"o}m}, \&
  {Primas}}]{Francois2007AA}
{Fran{\c c}ois} P. {et~al.}, 2007, \aap, 476, 935

\bibitem[{{Frebel} \& {Norris}(2015)}]{Frebel2015ARAA}
{Frebel} A., {Norris} J.~E., 2015, \araa, 53, 631

\bibitem[{{Freiburghaus}, {Rosswog} \& {Thielemann}(1999){Freiburghaus},
  {Rosswog}, \& {Thielemann}}]{freiburghaus1999ApJ}
{Freiburghaus} C., {Rosswog} S., {Thielemann} F.-K., 1999, \apjl, 525, L121

\bibitem[{{Ge} {et~al}\mbox{.}(2008){Ge}, {Mahadevan}, {Lee}, {Wan}, {Zhao},
  {van Eyken}, {Kane}, {Guo}, {Ford}, {Fleming}, {Crepp}, {Cohen}, {Groot},
  {Galvez}, {Liu}, {Agol}, {Gaudi}, {Ford}, {Schneider}, {Seager}, {Weinberg},
  \& {Eisenstein}}]{Ge2008ASPC}
{Ge} J. {et~al.}, 2008, in Astronomical Society of the Pacific Conference
  Series, Vol. 398, Extreme Solar Systems, {Fischer} D., {Rasio} F.~A.,
  {Thorsett} S.~E., {Wolszczan} A., eds., p. 449

\bibitem[{{Goriely}, {Bauswein} \& {Janka}(2011){Goriely}, {Bauswein}, \&
  {Janka}}]{Goriely2011ApJ}
{Goriely} S., {Bauswein} A., {Janka} H.-T., 2011, \apjl, 738, L32

\bibitem[{{Grevesse} \& {Sauval}(1998)}]{Grevesse1998}
{Grevesse} N., {Sauval} A.~J., 1998, \ssr, 85, 161

\bibitem[{{Gunn} {et~al}\mbox{.}(2006){Gunn}, {Siegmund}, {Mannery}, {Owen},
  {Hull}, {Leger}, {Carey}, {Knapp}, {York}, {Boroski}, {Kent}, {Lupton},
  {Rockosi}, {Evans}, {Waddell}, {Anderson}, {Annis}, {Barentine}, {Bartoszek},
  {Bastian}, {Bracker}, {Brewington}, {Briegel}, {Brinkmann}, {Brown}, {Carr},
  {Czarapata}, {Drennan}, {Dombeck}, {Federwitz}, {Gillespie}, {Gonzales},
  {Hansen}, {Harvanek}, {Hayes}, {Jordan}, {Kinney}, {Klaene}, {Kleinman},
  {Kron}, {Kresinski}, {Lee}, {Limmongkol}, {Lindenmeyer}, {Long}, {Loomis},
  {McGehee}, {Mantsch}, {Neilsen}, {Neswold}, {Newman}, {Nitta}, {Peoples},
  {Pier}, {Prieto}, {Prosapio}, {Rivetta}, {Schneider}, {Snedden}, \&
  {Wang}}]{Gunn2006AJ}
{Gunn} J.~E. {et~al.}, 2006, \aj, 131, 2332

\bibitem[{{Hansen}, {Montes} \& {Arcones}(2014){Hansen}, {Montes}, \&
  {Arcones}}]{Hansen-Montes2014ApJ}
{Hansen} C.~J., {Montes} F., {Arcones} A., 2014, \apj, 797, 123

\bibitem[{{Henden} {et~al}\mbox{.}(2015){Henden}, {Levine}, {Terrell}, \&
  {Welch}}]{Henden2015AAS}
{Henden} A.~A., {Levine} S., {Terrell} D., {Welch} D.~L., 2015, in American
  Astronomical Society Meeting Abstracts, Vol. 225, American Astronomical
  Society Meeting Abstracts, p. 336.16

\bibitem[{{Honda} {et~al}\mbox{.}(2004){Honda}, {Aoki}, {Kajino}, {Ando},
  {Beers}, {Izumiura}, {Sadakane}, \& {Takada-Hidai}}]{honda2004ApJ}
{Honda} S., {Aoki} W., {Kajino} T., {Ando} H., {Beers} T.~C., {Izumiura} H.,
  {Sadakane} K., {Takada-Hidai} M., 2004, \apj, 607, 474

\bibitem[{{Ishimaru}, {Wanajo} \& {Prantzos}(2015){Ishimaru}, {Wanajo}, \&
  {Prantzos}}]{Ishimaru_Wanajo_Prantzos_2015ApJ}
{Ishimaru} Y., {Wanajo} S., {Prantzos} N., 2015, \apjl, 804, L35

\bibitem[{{Ivans} {et~al}\mbox{.}(2003){Ivans}, {Sneden}, {James}, {Preston},
  {Fulbright}, {H{\"o}flich}, {Carney}, \& {Wheeler}}]{Ivans2003ApJ}
{Ivans} I.~I., {Sneden} C., {James} C.~R., {Preston} G.~W., {Fulbright} J.~P.,
  {H{\"o}flich} P.~A., {Carney} B.~W., {Wheeler} J.~C., 2003, \apj, 592, 906

\bibitem[{{King}(1997)}]{King1997AJ}
{King} J.~R., 1997, \aj, 113, 2302

\bibitem[{{Kirby} \& {Cohen}(2012)}]{Kirby2012AJ}
{Kirby} E.~N., {Cohen} J.~G., 2012, \aj, 144, 168

\bibitem[{{Kobayashi} {et~al}\mbox{.}(2014){Kobayashi}, {Ishigaki}, {Tominaga},
  \& {Nomoto}}]{Kobayashi2014ApJ}
{Kobayashi} C., {Ishigaki} M.~N., {Tominaga} N., {Nomoto} K., 2014, \apjl, 785,
  L5

\bibitem[{{Kobayashi} {et~al}\mbox{.}(2006){Kobayashi}, {Umeda}, {Nomoto},
  {Tominaga}, \& {Ohkubo}}]{Kobayashi2006ApJ}
{Kobayashi} C., {Umeda} H., {Nomoto} K., {Tominaga} N., {Ohkubo} T., 2006,
  \apj, 653, 1145

\bibitem[{{Kurucz}(2007)}]{K07}
{Kurucz} R.~L., 2007, Robert l. kurucz on-line database of observed and
  predicted atomic transitions

\bibitem[{{Kurucz}(2009)}]{K09}
{Kurucz} R.~L., 2009, Robert l. kurucz on-line database of observed and
  predicted atomic transitions

\bibitem[{{Kurucz}(2013)}]{K13}
{Kurucz} R.~L., 2013, Robert l. kurucz on-line database of observed and
  predicted atomic transitions

\bibitem[{{Lai} {et~al}\mbox{.}(2008){Lai}, {Bolte}, {Johnson}, {Lucatello},
  {Heger}, \& {Woosley}}]{Lai2008ApJ}
{Lai} D.~K., {Bolte} M., {Johnson} J.~A., {Lucatello} S., {Heger} A., {Woosley}
  S.~E., 2008, \apj, 681, 1524

\bibitem[{{Lai} {et~al}\mbox{.}(2009){Lai}, {Rockosi}, {Bolte}, {Johnson},
  {Beers}, {Lee}, {Allende Prieto}, \& {Yanny}}]{Lai2009ApJ}
{Lai} D.~K., {Rockosi} C.~M., {Bolte} M., {Johnson} J.~A., {Beers} T.~C., {Lee}
  Y.~S., {Allende Prieto} C., {Yanny} B., 2009, \apjl, 697, L63

\bibitem[{{Lattimer} {et~al}\mbox{.}(1977){Lattimer}, {Mackie}, {Ravenhall}, \&
  {Schramm}}]{lattimer1977ApJ}
{Lattimer} J.~M., {Mackie} F., {Ravenhall} D.~G., {Schramm} D.~N., 1977, \apj,
  213, 225

\bibitem[{{Lattimer} \& {Schramm}(1976)}]{lattimer_schramm1976ApJ}
{Lattimer} J.~M., {Schramm} D.~N., 1976, \apj, 210, 549

\bibitem[{{Lee} {et~al}\mbox{.}(2008){Lee}, {Beers}, {Sivarani}, {Johnson},
  {An}, {Wilhelm}, {Allende Prieto}, {Koesterke}, {Re Fiorentin},
  {Bailer-Jones}, {Norris}, {Yanny}, {Rockosi}, {Newberg}, {Cudworth}, \&
  {Pan}}]{LeeS2008AJ-SEGUE}
{Lee} Y.~S. {et~al.}, 2008, \aj, 136, 2050

\bibitem[{{Martin}, {Fuhr} \& {Wiese}(1988){Martin}, {Fuhr}, \& {Wiese}}]{MFW}
{Martin} G., {Fuhr} J., {Wiese} W., 1988, J. Phys. Chem. Ref. Data Suppl., 17

\bibitem[{{Mashonkina}, {Korn} \& {Przybilla}(2007){Mashonkina}, {Korn}, \&
  {Przybilla}}]{Mashonkina2007AA}
{Mashonkina} L., {Korn} A.~J., {Przybilla} N., 2007, \aap, 461, 261

\bibitem[{{McWilliam} {et~al}\mbox{.}(1995){McWilliam}, {Preston}, {Sneden}, \&
  {Searle}}]{McWilliam1995AJII}
{McWilliam} A., {Preston} G.~W., {Sneden} C., {Searle} L., 1995, \aj, 109, 2757

\bibitem[{{Meyer}(1989)}]{Meyer1989ApJ}
{Meyer} B.~S., 1989, \apj, 343, 254

\bibitem[{{Montes} {et~al}\mbox{.}(2007){Montes}, {Beers}, {Cowan}, {Elliot},
  {Farouqi}, {Gallino}, {Heil}, {Kratz}, {Pfeiffer}, {Pignatari}, \&
  {Schatz}}]{Montes2007ApJ}
{Montes} F. {et~al.}, 2007, \apj, 671, 1685

\bibitem[{{Otsuki} {et~al}\mbox{.}(2006){Otsuki}, {Honda}, {Aoki}, {Kajino}, \&
  {Mathews}}]{Otsuki2006ApJ}
{Otsuki} K., {Honda} S., {Aoki} W., {Kajino} T., {Mathews} G.~J., 2006, \apjl,
  641, L117

\bibitem[{{Plez}(2012)}]{Plez2012ascl}
{Plez} B., 2012, {Turbospectrum: Code for spectral synthesis}. Astrophysics
  Source Code Library

\bibitem[{{Prantzos}(2006)}]{prantzos2006astro.ph.12633P}
{Prantzos} N., 2006, ArXiv Astrophysics e-prints

\bibitem[{{Roederer} {et~al}\mbox{.}(2012){Roederer}, {Lawler}, {Sobeck},
  {Beers}, {Cowan}, {Frebel}, {Ivans}, {Schatz}, {Sneden}, \&
  {Thompson}}]{Roederer2012ApJS}
{Roederer} I.~U. {et~al.}, 2012, {New Hubble Space Telescope Observations of
  Heavy Elements in Four Metal-Poor Stars}

\bibitem[{{Roederer} {et~al}\mbox{.}(2014){Roederer}, {Preston}, {Thompson},
  {Shectman}, {Sneden}, {Burley}, \& {Kelson}}]{Roederer2014AJ}
{Roederer} I.~U., {Preston} G.~W., {Thompson} I.~B., {Shectman} S.~A., {Sneden}
  C., {Burley} G.~S., {Kelson} D.~D., 2014, \aj, 147, 136

\bibitem[{{Rosswog} {et~al}\mbox{.}(2014){Rosswog}, {Korobkin}, {Arcones},
  {Thielemann}, \& {Piran}}]{rosswog2014MNRAS}
{Rosswog} S., {Korobkin} O., {Arcones} A., {Thielemann} F.-K., {Piran} T.,
  2014, \mnras, 439, 744

\bibitem[{{Ryan} \& {Deliyannis}(1995)}]{Ryan1995ApJ}
{Ryan} S.~G., {Deliyannis} C.~P., 1995, \apj, 453, 819

\bibitem[{{Sbordone} {et~al}\mbox{.}(2010){Sbordone}, {Bonifacio}, {Caffau},
  {Ludwig}, {Behara}, {Gonz{\'a}lez Hern{\'a}ndez}, {Steffen}, {Cayrel},
  {Freytag}, {van't Veer}, {Molaro}, {Plez}, {Sivarani}, {Spite}, {Spite},
  {Beers}, {Christlieb}, {Fran{\c c}ois}, \& {Hill}}]{sbordone2010AA}
{Sbordone} L. {et~al.}, 2010, \aap, 522, A26

\bibitem[{{Schlegel}, {Finkbeiner} \& {Davis}(1998){Schlegel}, {Finkbeiner}, \&
  {Davis}}]{Schlegel1998ApJ}
{Schlegel} D.~J., {Finkbeiner} D.~P., {Davis} M., 1998, \apj, 500, 525

\bibitem[{{Smith} \& {Liszt}(1971)}]{SLa}
{Smith} W.~H., {Liszt} H.~S., 1971, Journal of the Optical Society of America
  (1917-1983), 61, 938, (SLa)

\bibitem[{{Sneden}, {Cowan} \& {Gallino}(2008){Sneden}, {Cowan}, \&
  {Gallino}}]{sneden2008ARAA}
{Sneden} C., {Cowan} J.~J., {Gallino} R., 2008, \araa, 46, 241

\bibitem[{{Sobeck}, {Lawler} \& {Sneden}(2007){Sobeck}, {Lawler}, \&
  {Sneden}}]{sobeck2007ApJ}
{Sobeck} J.~S., {Lawler} J.~E., {Sneden} C., 2007, \apj, 667, 1267

\bibitem[{{Spite} {et~al}\mbox{.}(2012){Spite}, {Andrievsky}, {Spite},
  {Caffau}, {Korotin}, {Bonifacio}, {Ludwig}, {Fran{\c c}ois}, \&
  {Cayrel}}]{Spite2012AA}
{Spite} M. {et~al.}, 2012, \aap, 541, A143

\bibitem[{{Spite} {et~al}\mbox{.}(2006){Spite}, {Cayrel}, {Hill}, {Spite},
  {Fran{\c c}ois}, {Plez}, {Bonifacio}, {Molaro}, {Depagne}, {Andersen},
  {Barbuy}, {Beers}, {Nordstr{\"o}m}, \& {Primas}}]{Spite2006A&A}
{Spite} M. {et~al.}, 2006, \aap, 455, 291

\bibitem[{{Suda} {et~al}\mbox{.}(2008){Suda}, {Katsuta}, {Yamada}, {Suwa},
  {Ishizuka}, {Komiya}, {Sorai}, {Aikawa}, \& {Fujimoto}}]{Suda2008PASJ}
{Suda} T. {et~al.}, 2008, \pasj, 60, 1159

\bibitem[{{Tafelmeyer} {et~al}\mbox{.}(2010){Tafelmeyer}, {Jablonka}, {Hill},
  {Shetrone}, {Tolstoy}, {Irwin}, {Battaglia}, {Helmi}, {Starkenburg}, {Venn},
  {Abel}, {Francois}, {Kaufer}, {North}, {Primas}, \&
  {Szeifert}}]{Tafelmeyer2010AA}
{Tafelmeyer} M. {et~al.}, 2010, \aap, 524, A58

\bibitem[{{Tanvir} {et~al}\mbox{.}(2013){Tanvir}, {Levan}, {Fruchter},
  {Hjorth}, {Hounsell}, {Wiersema}, \& {Tunnicliffe}}]{tanvir2013Nature}
{Tanvir} N.~R., {Levan} A.~J., {Fruchter} A.~S., {Hjorth} J., {Hounsell} R.~A.,
  {Wiersema} K., {Tunnicliffe} R.~L., 2013, \nat, 500, 547

\bibitem[{{Tsujimoto} \& {Shigeyama}(2014)}]{Tsujimoto_Shigeyama2014ApJ}
{Tsujimoto} T., {Shigeyama} T., 2014, \apjl, 795, L18

\bibitem[{{Vangioni} {et~al}\mbox{.}(2016){Vangioni}, {Goriely}, {Daigne},
  {Fran{\c c}ois}, \& {Belczynski}}]{vangioni2016MNRAS}
{Vangioni} E., {Goriely} S., {Daigne} F., {Fran{\c c}ois} P., {Belczynski} K.,
  2016, \mnras, 455, 17

\bibitem[{{Vargas} {et~al}\mbox{.}(2013){Vargas}, {Geha}, {Kirby}, \&
  {Simon}}]{vargas_2013ApJ}
{Vargas} L.~C., {Geha} M., {Kirby} E.~N., {Simon} J.~D., 2013, \apj, 767, 134

\bibitem[{{Wanajo} {et~al}\mbox{.}(2014){Wanajo}, {Sekiguchi}, {Nishimura},
  {Kiuchi}, {Kyutoku}, \& {Shibata}}]{wanajo2014ApJ}
{Wanajo} S., {Sekiguchi} Y., {Nishimura} N., {Kiuchi} K., {Kyutoku} K.,
  {Shibata} M., 2014, \apjl, 789, L39

\bibitem[{{Wiese}, {Smith} \& {Glennon}(1966){Wiese}, {Smith}, \&
  {Glennon}}]{WSG}
{Wiese} W.~L., {Smith} M.~W., {Glennon} B.~M., 1966, {Atomic transition
  probabilities. Vol.: Hydrogen through Neon. A critical data compilation},
  {Wiese, W.~L., Smith, M.~W., \& Glennon, B.~M.}, ed. US Government Printing
  Office, (WSG)

\bibitem[{{Wood} {et~al}\mbox{.}(2013){Wood}, {Lawler}, {Sneden}, \&
  {Cowan}}]{wood_lawler2013ApJS}
{Wood} M.~P., {Lawler} J.~E., {Sneden} C., {Cowan} J.~J., 2013, \apjs, 208, 27

\bibitem[{{Wood} {et~al}\mbox{.}(2014){Wood}, {Lawler}, {Sneden}, \&
  {Cowan}}]{wood2014ApJS}
{Wood} M.~P., {Lawler} J.~E., {Sneden} C., {Cowan} J.~J., 2014, \apjs, 211, 20

\bibitem[{{Wright} {et~al}\mbox{.}(2010){Wright}, {Eisenhardt}, {Mainzer},
  {Ressler}, {Cutri}, {Jarrett}, {Kirkpatrick}, {Padgett}, {McMillan},
  {Skrutskie}, {Stanford}, {Cohen}, {Walker}, {Mather}, {Leisawitz}, {Gautier},
  {McLean}, {Benford}, {Lonsdale}, {Blain}, {Mendez}, {Irace}, {Duval}, {Liu},
  {Royer}, {Heinrichsen}, {Howard}, {Shannon}, {Kendall}, {Walsh}, {Larsen},
  {Cardon}, {Schick}, {Schwalm}, {Abid}, {Fabinsky}, {Naes}, \&
  {Tsai}}]{Wright2010AJ}
{Wright} E.~L. {et~al.}, 2010, \aj, 140, 1868

\bibitem[{{Wright} \& {Howard}(2009)}]{Wright2009ApjS}
{Wright} J.~T., {Howard} A.~W., 2009, \apjs, 182, 205

\bibitem[{{Yong} {et~al}\mbox{.}(2013){Yong}, {Norris}, {Bessell},
  {Christlieb}, {Asplund}, {Beers}, {Barklem}, {Frebel}, \& {Ryan}}]{Yong2013a}
{Yong} D. {et~al.}, 2013, \apj, 762, 26

\end{thebibliography}
\newpage
\onecolumn
\appendix
\begin{center}
\section*{Appendix}
\end{center}
%-------------------------------------------
\begin{center}
{
\footnotesize
\begin{longtable}{r c c r r r r r r c c}
\caption{Atomic data and derived abundances for \gscc. Errors due to
$\delta$ T, $\delta$ log $\textit{g}$, and $ \delta \xi $
are added in quadrature} \\
\hline\hline
Element& $\lambda$ &  $\chi$ & log(gf)&  Equivalent & A(x) &abund         & abund           & abund          	& $\sigma$ & References  \\
       &        &         &        &   width     &      &$\delta$ T    & $\delta$ log $\textit{g}$  & $ \delta \xi 	$ &          &   \\
       & ($\rm \AA $)	&  (eV)   &        &    (m$ \rm \AA$)   &dex  &$\pm$ (250 K) &  $\pm$ (0.5)    & $\pm$0.2 km/s 	 & $\pm$     &    \\ 
\hline
\hline
\endfirsthead
\multicolumn{11}{c}{{\tablename\ \thetable{} -- continued from previous page}} \\ \hline \\
Element& $\lambda$   &  $\chi$  & log(gf) & Equivalent    & A(x)   & abund        & abund          & abund           & $\sigma$ & References    \\
       &             &          &         &       width   &        & $\delta$ T   & $\delta$ log g & $ \delta \xi$   &         &    \\
       &  $\rm \AA $ &  (eV)   &        & m$ \rm \AA$     &  dex   & $\pm$ 250 K  &  $\pm$ 0.5     &  $\pm$ 0.2 km/s &  $\pm$  &     \\     

\hline
\hline
\endhead
\hline
\endlastfoot

      Na  I & 5889.951  &  0.000 &   0.117 &  52.80 &   2.63 &  -0.199 &   0.022 &   0.019   &   0.201	&  (1)		\\	
      Na  I & 5895.924  &  0.000 &  -0.184 &  33.70 &   2.61 &  -0.193 &   0.009 &   0.009   &   0.193	&  (1)		\\
      Mg  I & 3829.355  &  2.710 &  -0.231 &  89.30 &   4.31 &  -0.209 &   0.104 &   0.058   &   0.241	&  (2)		\\
{$^*$}Mg  I & 3832.304  &  2.720 &   0.121 & 126.20 &   4.56 &  -0.268 &   0.210 &   0.051   &   0.344	&  (2) 		\\
{$^*$}Mg  I & 3838.290  &  2.720 &   0.415 & 151.00 &   4.51 &  -0.279 &   0.231 &   0.036   &   0.364	&  (3)	\\	
      Mg  I & 5172.684  &  2.710 &  -0.402 &  90.10 &   4.29 &  -0.226 &   0.108 &   0.046   &   0.255	&  (2)		\\
      Mg  I & 5183.604  &  2.720 &  -0.180 & 106.10 &   4.33 &  -0.250 &   0.155 &   0.052   &   0.299	&  (2)		\\
      Al  I & 3944.006  &  0.000 &  -0.623 &  48.20 &   2.58 &  -0.221 &   0.014 &   0.020   &   0.222	&  (4)		\\
      Al  I & 3961.520  &  0.010 &  -0.323 &  72.00 &   2.72 &  -0.235 &   0.056 &   0.045   &   0.246	&  (4)		\\
      Si  I & 3905.523  &  1.910 &  -0.744 &  92.10 &   4.11 &  -0.255 &   0.093 &   0.070   &   0.280	&  (5)		\\
%     Si  I & 4102.936  &  1.910 &  -2.828 &   7.10 &   4.24 &  -0.211 &  -0.032 &   0.002   &   0.213	&  (5)		\\
      Ca  I & 4226.728  &  0.000 &   0.357 & 104.70 &   3.04 &  -0.305 &   0.169 &   0.091   &   0.360	&  (5)		\\
      Ca  I & 4302.528  &  1.890 &   0.339 &  18.20 &   3.10 &  -0.155 &  -0.002 &   0.006   &   0.155	&  (5)		\\
{$^*$}Ca  I & 4434.957  &  1.890 &  -0.095 &  14.40 &   3.40 &  -0.154 &  -0.003 &   0.005   &   0.154	&  (5)		\\
      Ca  I & 4454.779  &  1.900 &   0.174 &  13.30 &   3.10 &  -0.154 &  -0.004 &   0.004   &   0.154	&  (5)		\\
      Ca  I & 6162.173  &  1.900 &  -0.043 &  13.90 &   3.26 &  -0.162 &   0.005 &   0.004   &   0.162	&  (5)		\\
      Sc II & 4246.822  &  0.310 &   0.305 &  40.00 &  -0.08 &  -0.153 &  -0.159 &   0.023   &   0.222	&  (6)		\\
      Ti II & 3757.685  &  1.570 &  -0.440 &  33.90 &   2.29 &  -0.096 &  -0.166 &   0.019   &   0.193	&  (7)	\\
      Ti II & 3759.292  &  0.610 &   0.280 & 105.10 &   2.23 &  -0.218 &   0.024 &   0.100   &   0.241	&  (7)	\\
      Ti II & 3761.321  &  0.570 &   0.180 &  75.00 &   1.59 &  -0.163 &  -0.088 &   0.104   &   0.212	&  (7)	\\
      Ti II & 3913.461  &  1.120 &  -0.360 &  46.70 &   1.99 &  -0.119 &  -0.156 &   0.035   &   0.199	&  (7)     \\
      Ti II & 4290.215  &  1.160 &  -0.870 &  14.10 &   1.73 &  -0.114 &  -0.169 &   0.006   &   0.204	&  (7)     \\
      Ti II & 4300.042  &  1.180 &  -0.460 &  21.40 &   1.56 &  -0.114 &  -0.168 &   0.010   &   0.203	&  (7)     \\
{$^*$}Ti II & 4395.839  &  1.240 &  -1.930 &  36.80 &   3.43 &  -0.115 &  -0.162 &   0.021   &   0.200	&  (7)     \\
      Ti II & 4443.801  &  1.080 &  -0.710 &  20.00 &   1.65 &  -0.119 &  -0.167 &   0.009   &   0.205	&  (7)     \\
      Ti II & 4468.493  &  1.130 &  -0.630 &  33.40 &   1.94 &  -0.118 &  -0.162 &   0.018   &   0.201	&  (7)     \\
      Ti II & 4501.270  &  1.120 &  -0.770 &  32.90 &   2.05 &  -0.120 &  -0.163 &   0.017   &   0.203	&  (7)     \\
      Ti II & 4533.969  &  1.240 &  -0.770 &  27.80 &   2.06 &  -0.114 &  -0.165 &   0.013   &   0.201	&  (8)          \\
      Ti II & 4549.622  &  1.580 &  -0.220 &  32.80 &   1.98 &  -0.101 &  -0.163 &   0.017   &   0.193	&  (7)     \\
      Cr  I & 4254.332  &  0.000 &  -0.090 &  52.30 &   2.24 &  -0.272 &   0.020 &   0.049   &   0.277	&  (9)	 \\
      Cr  I & 4274.796  &  0.000 &  -0.220 &  49.20 &   2.31 &  -0.270 &   0.016 &   0.043   &   0.274	&  (9)     \\
{$^*$}Cr  I & 4289.716  &  0.000 &  -0.370 &  17.40 &   1.73 &  -0.264 &   0.000 &   0.008   &   0.264	&  (9)     \\
      Mn  I & 4030.752  &  0.000 &  -0.431 &  37.30 &   1.65 &  -0.287 &   0.003 &   0.026   &   0.288	&  (5)          \\
      Mn  I & 4033.062  &  0.000 &  -0.590 &  39.30 &   1.85 &  -0.286 &   0.004 &   0.029   &   0.287	&  (5)          \\
      Fe  I & 3719.935  &  0.000 &  -0.431 & 113.30 &   3.89 &  -0.411 &   0.232 &   0.105   &   0.483	&  (5)          \\
      Fe  I & 3748.262  &  0.110 &  -0.961 &  92.30 &   4.08 &  -0.362 &   0.158 &   0.145   &   0.421	&  (5)          \\
      Fe  I & 3758.233  &  0.960 &   0.082 & 122.90 &   4.38 &  -0.371 &   0.234 &   0.076   &   0.445	&  (5)          \\
      Fe  I & 3763.789  &  0.990 &  -0.123 &  84.30 &   3.87 &  -0.308 &   0.120 &   0.130   &   0.355  	&  (5)		\\
      Fe  I & 3767.192  &  1.010 &  -0.279 &  78.20 &   3.88 &  -0.295 &   0.094 &   0.124   &   0.334 	&  (5)          \\
      Fe  I & 3795.002  &  0.990 &  -0.657 &  79.00 &   4.25 &  -0.296 &   0.096 &   0.126   &   0.336      &  (5)          \\
      Fe  I & 3815.840  &  1.490 &   0.303 & 103.60 &   4.32 &  -0.328 &   0.194 &   0.100   &   0.394  	&  (5)          \\
      Fe  I & 3820.425  &  0.860 &   0.214 & 122.50 &   4.12 &  -0.368 &   0.224 &   0.078   &   0.438	&  (5)          \\
      Fe  I & 3825.881  &  0.920 &   0.063 &  93.60 &   3.81 &  -0.327 &   0.152 &   0.125   &   0.382  	&  (5)          \\
      Fe  I & 3840.437  &  0.990 &  -0.400 &  82.40 &   4.07 &  -0.302 &   0.107 &   0.127   &   0.345  	&  (5)          \\
      Fe  I & 3841.048  &  1.610 &   0.008 &  74.10 &   4.04 &  -0.267 &   0.084 &   0.108   &   0.300  	&  (5)          \\
      Fe  I & 3849.966  &  1.010 &  -0.764 &  69.40 &   4.09 &  -0.280 &   0.056 &   0.104   &   0.304  	&  (5)          \\
      Fe  I & 3856.371  &  0.050 &  -1.236 &  86.20 &   4.09 &  -0.349 &   0.122 &   0.147   &   0.398	&  (5)          \\
      Fe  I & 3859.911  &  0.000 &  -0.658 & 111.70 &   4.05 &  -0.407 &   0.223 &   0.113   &   0.478  &  (5)          \\
      Fe  I & 3865.523  &  1.010 &  -0.904 &  64.40 &   4.09 &  -0.274 &   0.039 &   0.090   &   0.291  &  (5)          \\
      Fe  I & 3872.501  &  0.990 &  -0.928 &  77.60 &   4.46 &  -0.294 &   0.086 &   0.121   &   0.329  &  (5)          \\
      Fe  I & 3886.282  &  0.050 &  -1.010 &  95.00 &   4.09 &  -0.371 &   0.162 &   0.144   &   0.430  &  (5)          \\
      Fe  I & 3887.048  &  0.920 &  -1.118 &  44.40 &   3.73 &  -0.267 &   0.003 &   0.038   &   0.270	& (5)          \\
      Fe  I & 3895.656  &  0.110 &  -1.606 &  88.40 &   4.57 &  -0.352 &   0.133 &   0.147   &   0.404  &  (5)          \\
      Fe  I & 3899.707  &  0.090 &  -1.461 &  83.10 &   4.25 &  -0.340 &   0.106 &   0.144   &   0.384	& (5)          \\
      Fe  I & 3920.258  &  0.120 &  -1.678 &  63.90 &   3.94 &  -0.309 &   0.033 &   0.094   &   0.325  &  (5)          \\
      Fe  I & 3922.912  &  0.050 &  -1.597 &  79.60 &   4.24 &  -0.334 &   0.089 &   0.139   &   0.373  &  (5)          \\
      Fe  I & 3927.920  &  0.110 &  -1.473 &  69.70 &   3.88 &  -0.316 &   0.050 &   0.113   &   0.339	&  (5)		\\
      Fe  I & 4005.242  &  1.560 &  -0.543 &  55.00 &   4.03 &  -0.247 &   0.022 &   0.060   &   0.255  &  (5)          \\
      Fe  I & 4045.812  &  1.490 &   0.334 &  93.70 &   4.04 &  -0.311 &   0.158 &   0.118   &   0.368  &  (5)          \\
      Fe  I & 4063.594  &  1.560 &   0.122 &  91.50 &   4.26 &  -0.305 &   0.150 &   0.118   &   0.360  &  (5)          \\
      Fe  I & 4132.058  &  1.610 &  -0.565 &  58.10 &   4.15 &  -0.248 &   0.029 &   0.066   &   0.258  &  (5)          \\
      Fe  I & 4143.868  &  1.560 &  -0.431 &  64.60 &   4.12 &  -0.257 &   0.044 &   0.083   &   0.274	&  (5)          \\
      Fe  I & 4187.039  &  2.450 &  -0.445 &  20.60 &   4.04 &  -0.201 &  -0.004 &   0.009   &   0.201  &  (5)          \\
      Fe  I & 4187.795  &  2.430 &  -0.448 &  23.40 &   4.09 &  -0.204 &  -0.002 &   0.011   &   0.204  &  (5)          \\
      Fe  I & 4198.304  &  2.400 &  -0.614 &  23.00 &   4.22 &  -0.204 &  -0.002 &   0.011   &   0.204  &  (5)          \\
      Fe  I & 4202.029  &  1.490 &  -0.493 &  48.40 &   3.73 &  -0.247 &   0.011 &   0.043   &   0.251  &  (5)          \\
      Fe  I & 4227.427  &  3.330 &   0.465 &  26.60 &   4.17 &  -0.166 &  -0.003 &   0.014   &   0.167	&  (5)          \\
      Fe  I & 4250.119  &  2.470 &  -0.302 &  29.00 &   4.12 &  -0.203 &   0.003 &   0.015   &   0.204  &  (5)          \\
      Fe  I & 4250.787  &  1.560 &  -0.590 &  51.60 &   3.96 &  -0.247 &   0.014 &   0.049   &   0.252  &  (5)          \\
      Fe  I & 4260.474  &  2.400 &   0.146 &  47.00 &   3.96 &  -0.217 &   0.028 &   0.034   &   0.221  &  (5)          \\
      Fe  I & 4271.760  &  1.490 &   0.150 &  72.60 &   3.65 &  -0.271 &   0.069 &   0.102   &   0.298  &  (5)          \\
      Fe  I & 4282.403  &  2.180 &  -0.693 &  30.20 &   4.24 &  -0.213 &  -0.008 &   0.018   &   0.214  &  (5)          \\
      Fe  I & 4325.762  &  1.610 &   0.114 &  62.40 &   3.55 &  -0.255 &   0.038 &   0.075   &   0.269  &  (5)          \\
      Fe  I & 4375.930  &  0.000 &  -3.150 &  25.50 &   4.33 &  -0.305 &  -0.003 &   0.014   &   0.305  &  (5)          \\
      Fe  I & 4383.545  &  1.490 &   0.100 &  87.50 &   4.06 &  -0.300 &   0.127 &   0.120   &   0.347  &  (5)          \\
      Fe  I & 4404.750  &  1.560 &  -0.346 &  68.20 &   4.09 &  -0.263 &   0.055 &   0.090   &   0.283  &  (5)		\\
      Fe  I & 4415.122  &  1.610 &  -0.893 &  44.20 &   4.14 &  -0.242 &   0.007 &   0.035   &   0.245  &  (5)          \\
      Fe  I & 4427.310  &  0.050 &  -3.175 &  31.90 &   4.55 &  -0.304 &   0.000 &   0.020   &   0.305  &  (5)          \\
      Fe  I & 4528.614  &  2.180 &  -0.850 &  21.60 &   4.16 &  -0.214 &  -0.005 &   0.010   &   0.214  &  (5)          \\
      Fe  I & 4890.755  &  2.880 &  -0.278 &  18.70 &   4.20 &  -0.188 &  -0.002 &   0.008   &   0.188  &  (5)          \\
      Fe  I & 4891.492  &  2.850 &   0.013 &  21.70 &   3.97 &  -0.189 &   0.000 &   0.010   &   0.189  &  (5)          \\
      Fe  I & 4920.502  &  2.830 &   0.000 &  34.90 &   4.26 &  -0.197 &   0.013 &   0.018   &   0.198  & (5)          \\
      Fe  I & 5171.596  &  1.490 &  -1.497 &  19.50 &   4.00 &  -0.246 &  -0.003 &   0.009   &   0.246  &  (5)          \\
      Fe  I & 5227.190  &  1.560 &  -1.062 &  31.40 &   3.93 &  -0.245 &   0.003 &   0.019   &   0.246  &  (5)          \\
      Fe  I & 5269.537  &  0.860 &  -1.075 &  60.80 &   3.83 &  -0.291 &   0.033 &   0.067   &   0.300  &  (5)          \\
      Fe  I & 5270.356  &  1.610 &  -1.200 &  30.00 &   4.09 &  -0.243 &   0.002 &   0.017   &   0.244  &  (5)          \\
      Fe  I & 5328.039  &  0.920 &  -1.236 &  53.90 &   3.90 &  -0.283 &   0.022 &   0.051   &   0.288  &  (5)          \\
      Fe  I & 5328.531  &  1.560 &  -1.718 &  19.70 &   4.29 &  -0.244 &  -0.003 &   0.009   &   0.244  &  (5)          \\
      Fe  I & 5371.490  &  0.960 &  -1.418 &  47.70 &   3.99 &  -0.278 &   0.014 &   0.039   &   0.281  &  (5)          \\
      Fe  I & 5397.128  &  0.920 &  -1.772 &  34.20 &   4.02 &  -0.274 &   0.004 &   0.021   &   0.275  &  (5)          \\
      Fe  I & 5405.775  &  0.990 &  -1.633 &  38.20 &   4.04 &  -0.273 &   0.006 &   0.025   &   0.274  &  (5)          \\
      Fe  I & 5429.696  &  0.960 &  -1.655 &  37.90 &   4.02 &  -0.274 &   0.006 &   0.024   &   0.275  &  (5)          \\
      Fe  I & 5434.524  &  1.010 &  -1.907 &  27.20 &   4.10 &  -0.269 &   0.001 &   0.014   &   0.269  &  (5)          \\
      Fe  I & 5446.916  &  0.990 &  -1.698 &  31.50 &   3.96 &  -0.271 &   0.003 &   0.018   &   0.272  &  (5)          \\
      Fe II & 4233.162  &  2.580 &  -1.884 &  21.50 &   4.34 &  -0.049 &  -0.181 &   0.011   &   0.188  &  (10)		\\
      Fe II & 4923.921  &  2.890 &  -1.559 &  23.50 &   4.33 &  -0.041 &  -0.178 &   0.012   &   0.183  &  (10)          \\
      Fe II & 5018.440  &  2.890 &  -1.399 &  39.20 &   4.52 &  -0.044 &  -0.172 &   0.026   &   0.179  &  (10)          \\
      Fe II & 5169.028  &  2.890 &  -1.300 &  39.20 &   4.41 &  -0.045 &  -0.170 &   0.026   &   0.178  &  (10)          \\
      Ni  I & 3783.530  &  0.422 &  -1.400 &  41.70 &   3.16 &  -0.274 &  -0.009 &   0.036   &   0.277  &  (11)     \\
      Ni  I & 3807.144  &  0.422 &  -1.230 &  54.00 &   3.27 &  -0.279 &   0.009 &   0.064   &   0.286  &  (11)     \\
      Ni  I & 3858.297  &  0.422 &  -0.960 &  55.80 &   3.03 &  -0.281 &   0.012 &   0.069   &   0.290  &  (11)     \\
      Sr II & 4077.709  &  0.000 &   0.150 &  56.60 &  -0.79 &  -0.175 &  -0.135 &   0.087   &   0.238  &  (12)     \\
      Sr II & 4215.519  &  0.000 &  -0.170 &  33.70 &  -1.04 &  -0.165 &  -0.161 &   0.027   &   0.232  &  (12)     \\

\label{atomic}                                                                                                          
\end{longtable}                                                                                                        
$^{*}$ Refers to those lines which are not used for calculating the mean 
abundance of the respective elements. \\
(1) \citet{WSG}, (2) \citet{AZS}, (3) NIST 2$^\dagger$, (4) \citet{SLa}, (5) \citet{K07}, (6) \citet{K09}, 
(7) \citet{wood_lawler2013ApJS}, (8) \citet{MFW},(9) \citet{sobeck2007ApJ}, (10) \citet{K13},(11) \citet{wood2014ApJS}, 
(12) \citet{Roederer2012ApJS}(and references therein).\\
$^\dagger$ Martin, W. C., Fuhr, J. R., \& Kelleher, D. E. et al. 2002, NIST Atomic Spectra Database (Gaithersburg, MD: NIST), Version 2.0
}                                                 
\end{center}
\end{document}